\documentclass[notitlepage,nofootinbib,preprintnumbers,amssymb,superscriptaddress]{revtex4-2}
\usepackage{amsfonts,amsmath,amssymb,mathtools,graphicx,color,bm}
\definecolor{ultramarine}{rgb}{0.07, 0.04, 0.56}
\definecolor{cadmiumgreen}{rgb}{0.0, 0.42, 0.24}
\definecolor{greenmasato}{rgb}{0.0, 0.70, 0.00}
\definecolor{indigo(dye)}{rgb}{0.0, 0.25, 0.42}
\usepackage[linktocpage=true]{hyperref}
\hypersetup{
colorlinks=true,
citecolor=ultramarine,
linkcolor=cadmiumgreen,
urlcolor=indigo(dye),
}

\usepackage{autobreak}
\newcommand{\D}{{\rm d}}

\newcommand{\fr}[2]{\frac{#1}{#2}}
\newcommand{\pa}{\partial}
\newcommand{\ti}{\tilde}
\newcommand{\na}{\nabla}
\newcommand{\bra}[1]{\left( #1 \right)}
\newcommand{\brb}[1]{\left[ #1 \right]}
\newcommand{\brc}[1]{\left\{ #1 \right\}}

\newcommand{\be}{\begin{equation}}  
\newcommand{\ee}{\end{equation}}
\newcommand{\bem}{\begin{pmatrix}}
\newcommand{\eem}{\end{pmatrix}}
\newcommand{\Mpl}{M_{\rm Pl}}

\newcommand{\ep}{\epsilon}
\newcommand{\vae}{\varepsilon}

\newcommand{\mn}{{\mu \nu}}

\newcommand{\mC}{\mathcal{C}}

\newcommand{\mK}{\mathcal{K}}

\newcommand{\mO}{\mathcal{O}}
\newcommand{\mP}{\mathcal{P}}
\newcommand{\mQ}{\mathcal{Q}}
\newcommand{\mV}{\mathcal{V}}
\newcommand{\mW}{\mathcal{W}}
\newcommand{\sizecorr}[1]{\makebox[0cm]{\phantom{$\displaystyle #1$}}}

\begin{document}

\preprint{RUP-23-26, YITP-23-147}

\title{Exact solution for rotating black holes in parity-violating gravity}

\author{Hiroaki W.~H.~Tahara}
\affiliation{Department of Physics, Rikkyo University, Toshima, Tokyo 171-8501, Japan}

\author{Kazufumi Takahashi}
\affiliation{Center for Gravitational Physics and Quantum Information, Yukawa Institute for Theoretical Physics, Kyoto University, 606-8502, Kyoto, Japan}

\author{Masato Minamitsuji}
\affiliation{Centro de Astrof\'{\i}sica e Gravita\c c\~ao  - CENTRA, Departamento de F\'{\i}sica, Instituto Superior T\'ecnico - IST,
Universidade de Lisboa - UL, Av.~Rovisco Pais 1, 1049-001 Lisboa, Portugal}

\author{Hayato Motohashi}
\affiliation{Division of Liberal Arts, Kogakuin University, 2665-1 Nakano-machi, Hachioji, Tokyo 192-0015, Japan}

\begin{abstract}
It has recently been pointed out that one can construct invertible conformal transformations with a parity-violating conformal factor, which can be employed to generate a novel class of parity-violating ghost-free metric theories from general relativity.
We obtain exact solutions for rotating black holes in such theories by performing the conformal transformation on the Kerr solution in general relativity, which we dub {\it conformal Kerr} solutions.
We explore the geodesic motion of a test particle in the conformal Kerr spacetime.
While null geodesics remain the same as those in the Kerr spacetime, timelike geodesics exhibit interesting differences due to an effective external force caused by the parity-violating conformal factor.
\end{abstract}

\maketitle

%%%%%%%%%%%%%%%%%%%%%%%%%%%%%%%%%%%%%%%%%%%%%%%%%%%%%%%%%%%%%%%%%%%%%%%%%%%%%%%%%%%%
%%%%%%%%%%%%%%%%%%%%%%%%%%%%%%%%%%%%%%%%%%%%%%%%%%%%%%%%%%%%%%%%%%%%%%%%%%%%%%%%%%%%
%	Introduction
%%%%%%%%%%%%%%%%%%%%%%%%%%%%%%%%%%%%%%%%%%%%%%%%%%%%%%%%%%%%%%%%%%%%%%%%%%%%%%%%%%%%
%%%%%%%%%%%%%%%%%%%%%%%%%%%%%%%%%%%%%%%%%%%%%%%%%%%%%%%%%%%%%%%%%%%%%%%%%%%%%%%%%%%%
\section{Introduction}\label{sec:intro}

Symmetries play essential roles in physics.
For instance, the Lorentz symmetry is one of the fundamental symmetries in particle physics and general relativity (GR).
In the Standard Model of particle physics, invariance under charge conjugation, parity transformation, and time reversal (CPT symmetry) is another fundamental symmetry.
While the Standard Model respects the CPT symmetry, each individual symmetry may be broken.
In fact, in 1950s, it was found that the parity conservation is broken in the weak sector~\cite{Lee:1956qn,Wu:1957my}, which is the only parity violation observed in nature so far.
The violation of the CP symmetry in the weak sector was also detected~\cite{Christenson:1964fg,NA48:1999szy}.
The parity violation may also occur in a new particle sector beyond the Standard Model or gravity sector, which may be detected via cosmological and astrophysical probes.
The possibility to detect the parity violation in electromagnetism and/or gravity sector on the cosmic microwave background has been attracting much attention for a long time~\cite{Lue:1998mq,Alexander:2004us,Li:2006ss,Satoh:2007gn,Saito:2007kt,Soda:2011am,Sorbo:2011rz,Kaufman:2014rpa,Masui:2017fzw,Bartolo:2017szm,Bartolo:2018elp,Minami:2020odp,Wang:2021gqm,Komatsu:2022nvu,Caravano:2022epk,Figueroa:2023oxc}.
More recently, advances in observations of gravitational waves have also enabled us to detect the possible parity violation in our Universe~\cite{Seto:2007tn,Yunes:2010yf,Cook:2011hg,Yagi:2017zhb,Alexander:2017jmt,Nishizawa:2018srh,Zhao:2019xmm,Qiao:2019hkz,Yamada:2020zvt,Perkins:2020tra,Perkins:2021mhb,Gong:2021jgg,Okounkova:2021xjv,Qiao:2022mln,Jenks:2023pmk}.

From a theoretical point of view, we can consider parity-violating theories of gravity, for instance by incorporating the Chern-Simons term (or Pontryagin density) in the action, which is known as Chern-Simons gravity~\cite{Jackiw:2003pm,Smith:2007jm} (see also Ref.~\cite{Alexander:2009tp} for a review).
In such a parity-violating theory of gravity, spherically symmetric solutions are not modified from those in GR, since the Pontryagin density trivially vanishes under the spherical symmetry.
Therefore, in order to observe the parity-violation in a spherically symmetric black hole (BH) background, one needs to consider perturbations about it~\cite{Yunes:2007ss,Cardoso:2009pk,Molina:2010fb,Garfinkle:2010zx,Pani:2011xj,Motohashi:2011pw,Motohashi:2011ds,Kimura:2018nxk,Macedo:2018txb,Wagle:2021tam}.
To the best of our knowledge, all the parity-violating gravitational theories studied so far, including Chern-Simons gravity, yield ghost degrees of freedom in general, which was confirmed for perturbations on a static and spherically symmetric background~\cite{Motohashi:2011pw,Motohashi:2011ds}. 
Hence, they should be regarded as an effective field theory which is valid at most up to an energy scale below the mass of the ghosts.

Another possibility to detect the parity violation in the gravity sector is to consider an axisymmetric spacetime or rotating BH, where the effects of parity violation can show up even at the background level.
In particular, the faster the BH rotates, the more significant the impact of the parity violation is expected to be.
Rotating BHs in Chern-Simons gravity have been studied perturbatively in the slow-rotation~\cite{Konno:2007ze,Konno:2009kg,Yunes:2009hc,Pani:2011gy,Yagi:2012ya,Maselli:2017kic,Srivastava:2021imr} and near-extremal~\cite{McNees:2015srl} approximations, as well as numerically~\cite{Stein:2014xba,Delsate:2018ome,Cunha:2018uzc} (see also Refs.~\cite{Berti:2015itd,Berti:2018cxi}). 
In dynamical Chern-Simons gravity, configurations of a scalar field around rotating BHs developed via spontaneous scalarization and superradiant instabilities have been investigated in Refs.~\cite{Myung:2021fzo,Doneva:2021dcc,Chatzifotis:2022mob,Zhang:2022sgt,Lin:2023npr} and Ref.~\cite{Alexander:2022avt}, respectively.
Also, observational signatures of Chern-Simons gravity have been investigated in Refs.~\cite{Smith:2007jm,Yagi:2011xp,Yagi:2012vf,Stein:2013wza,Yunes:2016jcc,Okounkova:2017yby}.
On the other hand, it was argued in Ref.~\cite{Grumiller:2007rv} that physical rotating BHs can exist in nondynamical Chern-Simons gravity only if the metric breaks the stationarity, axisymmetry, or energy-momentum conservation.
Slowly rotating BHs in other parity-violating theories with higher-curvature corrections and observational signatures from them have also been studied~\cite{Cardoso:2018ptl,Cano:2019ore,Sennett:2019bpc,Nakashi:2020phm,Cano:2020cao,Cano:2021myl,Cano:2022wwo}.
An effective field theory for perturbations of axisymmetric and slowly rotating solutions was discussed including the parity-violating terms in Ref.~\cite{Hui:2021cpm}.
A phenomenological parametrization for parity-violating Kerr-like metrics has been studied in Refs.~\cite{Chen:2020aix,Chen:2021ryb} within a family of metrics where the geodesic equations are separable~\cite{Papadopoulos:2018nvd}.
However, despite such extensive studies of BHs in parity-violating theories, no exact solution of rotating BHs has been reported so far.

A possible way to generate such nontrivial solutions is to perform an invertible conformal/disformal transformation on known solutions in GR or any other modified gravity models.
For instance, the disformal transformation of static and spherically symmetric solutions with a stealth scalar profile has been studied in Refs.~\cite{Takahashi:2019oxz,BenAchour:2020wiw}.
Other known examples are a disformal Kerr solution~\cite{Anson:2020trg,BenAchour:2020fgy} and a conformal Kerr solution~\cite{Babichev:2023mgk},\footnote{In \S\ref{sec:conformal_Kerr}, we shall also refer to our BH solution as the conformal Kerr solution, though it is qualitatively different from that in Ref.~\cite{Babichev:2023mgk}:
Their solution does not correspond to a stationary metric, where an explicit time dependence is introduced by the conformal transformation.} which are generated from the stealth Kerr solution in scalar-tensor theories.
The point is that the invertibility of the transformation guarantees the existence of a gravitational theory that can accommodate the generated metric as an exact solution.
We also note that an invertible transformation maps a gravitational theory to another theory without changing the number of propagating degrees of freedom~\cite{Domenech:2015tca,Takahashi:2017zgr}.
In particular, if the seed theory is GR, then the generated theory possesses two tensorial degrees of freedom and is free from the problem of Ostrogradsky ghosts~\cite{Woodard:2015zca,Motohashi:2014opa,Motohashi:2020psc,Aoki:2020gfv} just as in GR.
Recently, the authors of Ref.~\cite{Takahashi:2022mew} found a new class of invertible conformal transformations involving a parity-violating interaction.
By performing this novel invertible transformation on the Kerr solution in GR, one can obtain a rotating BH metric as an exact solution in a parity-violating ghost-free theory.
The aim of the present paper is to demonstrate the strategy above and investigate the properties of the solution, including the geodesic motion of a test particle.

The rest of this paper is organized as follows.
In \S\ref{sec:trs}, we review the invertible conformal transformation involving a parity-violating interaction proposed in Ref.~\cite{Takahashi:2022mew} and use it to construct a rotating BH solution in a parity-violating gravitational theory.
In \S\ref{sec:geodesics}, we study the geodesic motion of a test particle to clarify how it is different from the one in GR.
Finally, we draw our conclusions in \S\ref{sec:conc}.

%%%%%%%%%%%%%%%%%%%%%%%%%%%%%%%%%%%%%%%%%%%%%%%%%%%%%%%%%%%%%%%%%%%%%%%%%%%%%%%%%%%%
%%%%%%%%%%%%%%%%%%%%%%%%%%%%%%%%%%%%%%%%%%%%%%%%%%%%%%%%%%%%%%%%%%%%%%%%%%%%%%%%%%%%
%	Black holes in parity-violating gravity from conformal transformation
%%%%%%%%%%%%%%%%%%%%%%%%%%%%%%%%%%%%%%%%%%%%%%%%%%%%%%%%%%%%%%%%%%%%%%%%%%%%%%%%%%%%
%%%%%%%%%%%%%%%%%%%%%%%%%%%%%%%%%%%%%%%%%%%%%%%%%%%%%%%%%%%%%%%%%%%%%%%%%%%%%%%%%%%%
\section{Black holes in parity-violating gravity from conformal transformation}
\label{sec:trs}

\subsection{Transformation law}
\label{sec:trnsf_law}

Let us briefly review the invertible conformal transformation involving curvature invariants introduced in Ref.~\cite{Takahashi:2022mew}.
We consider the conformal transformation 
    \be
    \bar{g}_\mn[g]=\Omega(\mC,\mP)g_\mn, \label{invertible_tr}
    \ee
where $\Omega$ is a positive definite function of $\mC$ and $\mP$, with $\mC$ being a contraction of the Weyl tensor~$\mW^\mu{}_{\nu\lambda\sigma}$ and $\mP$ being the Chern-Simons (or Pontryagin) term defined by~\cite{Jackiw:2003pm,Grumiller:2007rv}
    \be
    \begin{split} 
    \mC&\coloneqq \mW^{\alpha\beta\gamma\delta}\mW_{\alpha\beta\gamma\delta}
    =R^{\alpha\beta\gamma\delta}R_{\alpha\beta\gamma\delta}-2R^{\alpha\beta}R_{\alpha\beta}+\fr{1}{3}R^2, \\
    \mP&\coloneqq \fr{1}{2}\vae^{\alpha\beta\gamma\delta}R^{\mu\nu}{}_{\alpha\beta}R_{\mu\nu\gamma\delta}
    = \fr{1}{2}\vae^{\alpha\beta\gamma\delta}\mW^{\mu\nu}{}_{\alpha\beta}\mW_{\mu\nu\gamma\delta}.
    \end{split}\label{CP}
    \ee
The totally antisymmetric tensor~$\vae^{\alpha\beta\gamma\delta}$ in the definition of $\mP$ manifests its parity-violating nature.
It should be noted that $\mC$ and $\mP$ are the only scalar quantities that are ``conformally covariant'' (i.e., invariant under a conformal transformation up to some powers of the conformal factor) up to the quadratic order in the curvature tensor.
Indeed, they are covariant with weight~$-2$ under any conformal transformation:
    \be
    \bar{\mC}=\Omega^{-2}\mC, \qquad
    \bar{\mP}=\Omega^{-2}\mP. \label{tr_CP}
    \ee
Note that $\Omega$ here can be identified with the one in \eqref{invertible_tr}.
In this case, Eq.~\eqref{tr_CP} defines a map~$(\mC,\mP)\mapsto (\bar{\mC},\bar{\mP})$, which can be solved for $\mC$ and $\mP$ at least locally if the Jacobian determinant is nonvanishing, i.e.,
	\be
	\left|\fr{\pa(\bar{\mC},\bar{\mP})}{\pa(\mC,\mP)}\right|
	\propto 1-\fr{2\mC}{\Omega}\fr{\pa\Omega}{\pa\mC}-\fr{2\mP}{\Omega}\fr{\pa\Omega}{\pa\mP}\ne 0. \label{invertible_cond}
	\ee
Once we obtain $\mC$ and $\mP$ as functions of $\bar{\mC}$ and $\bar{\mP}$, the inverse transformation of \eqref{invertible_tr} can be given by
    \be
    g_\mn[\bar{g}]=\bar{\Omega}(\bar{\mC},\bar{\mP})\bar{g}_\mn, \qquad
    \bar{\Omega}(\bar{\mC},\bar{\mP})\coloneqq \Omega(\mC,\mP)^{-1}. \label{g_to_gbar}
    \ee

More generally, 
one could include any other conformally covariant scalar quantities (e.g., $\mW^{\alpha\beta}{}_{\gamma\delta}\mW^{\gamma\delta}{}_{\mu\nu}\mW^{\mu\nu}{}_{\alpha\beta}$) in the conformal factor to generalize the transformation law~\eqref{invertible_tr}.
When the conformal factor is a function of conformally covariant quantities~$\mC_I$ ($I=1,2,\cdots$) such that $\bar{\mC}_I=\Omega^{w_I}\mC_I$, the invertibility condition is given by
    \be
    \left|\fr{\pa\bar{\mC}_I}{\pa\mC_J}\right|
	\propto 1+\sum_I\fr{w_I\mC_I}{\Omega}\fr{\pa\Omega}{\pa\mC_I}\ne 0.
    \ee

By use of such invertible transformations, one can generate a class of ghost-free higher-derivative metric theories from GR since an invertible transformation does not change the number of physical degrees of freedom~\cite{Domenech:2015tca,Takahashi:2017zgr}.
In other words, after performing the invertible conformal transformation~\eqref{invertible_tr} or its generalization mentioned above, the higher-order derivatives of the metric automatically satisfy the degeneracy condition~\cite{Langlois:2015cwa,Motohashi:2016ftl}.\footnote{Noninvertible transformations could also be used to generate theories with degenerate higher-derivative terms~\cite{Chamseddine:2013kea,Takahashi:2017pje,Langlois:2018jdg}. Having said that, we do not consider this possibility in the present paper.}
Starting from the Einstein-Hilbert action
    \be
    S_{\rm EH}[{g}]=\int \D^4x\sqrt{-g}\,\fr{\Mpl^2}{2}{R},
    \ee
we obtain the following action:
    \be
    S[\bar{g}]\coloneqq S_{\rm EH}[g[\bar{g}]]
    =\int \D^4x\sqrt{-\bar{g}}\,\fr{\Mpl^2}{2}\bra{\bar{\Omega}\bar{R}+\fr{3}{2\bar{\Omega}}\bar{\na}^\lambda\bar{\Omega}\bar{\na}_\lambda\bar{\Omega}},
    \label{conformal_EH}
    \ee
with $\Mpl$ being the reduced Planck mass.
Here, we have omitted boundary terms, and $\bar g_{\mu\nu}$ is now regarded as the physical metric.
Note that the action~\eqref{conformal_EH} involves the parity-violating Chern-Simons term through $\bar{\Omega}$.
It should also be noted that the action is written only by the metric, unlike the case of Chern-Simons gravity where a scalar field (which can be either dynamical or nondynamical) is present.

As a concrete example, let us consider the case where the conformal factor has the form\footnote{Another possible choice of the conformal factor would be $\Omega=\exp(\alpha\mP/\Lambda^4)$, but in this case the invertibility condition~\eqref{invertible_cond} is violated at $\mP=\Lambda^4/2\alpha$.}
    \be
    \Omega=1+\tanh\bra{\fr{\alpha\mP}{\Lambda^4}},
    \label{Omega_P}
    \ee
where $\alpha$ is a dimensionless parameter of order unity and $\Lambda$ is some mass scale.
The relation between $\mP$ and $\bar{\mP}$ is given by
    \begin{align}
    \bar{\mP}=\brb{1+\tanh\bra{\fr{\alpha\mP}{\Lambda^4}}}^{-2}\mP,
    \label{P2Pbar}
    \end{align}
which satisfies $\D\bar{\mP}/\D\mP>0$ for any finite $\mP$, and hence the transformation is always invertible.
The series expansion of $\mP(\bar{\mP})$ around $\bar{\mP}=0$ is formally given by
    \be
    \mP(\bar{\mP})=\sum_{n=1}^\infty c_n\bar{\mP}^n, \qquad
    c_n\coloneqq \fr{1}{n!}\lim_{\mP\to0}\fr{\D^{n-1}}{\D\mP^{n-1}}\brc{\brb{1+\tanh\bra{\fr{\alpha\mP}{\Lambda^4}}}^{2n}},
    \ee
or written explicitly, 
    \be
    \mP=\bar{\mP}\brb{1+2\bra{\fr{\alpha\bar{\mP}}{\Lambda^4}}+5\bra{\fr{\alpha\bar{\mP}}{\Lambda^4}}^2+\fr{40}{3}\bra{\fr{\alpha\bar{\mP}}{\Lambda^4}}^3+36\bra{\fr{\alpha\bar{\mP}}{\Lambda^4}}^4+\fr{478}{5}\bra{\fr{\alpha\bar{\mP}}{\Lambda^4}}^5+\fr{10946}{45}\bra{\fr{\alpha\bar{\mP}}{\Lambda^4}}^6+\cdots}.
    \ee
For the above choice of the conformal factor, the action~\eqref{conformal_EH} takes the form
    \be
    S[\bar{g}]
    =\int \D^4x\sqrt{-\bar{g}}\,\fr{\Mpl^2}{2}\brb{1+\tanh\bra{\fr{\alpha\mP}{\Lambda^4}}}^{-1}\brc{\bar{R}+\fr{3\alpha^2}{2\Lambda^8}\brb{\fr{\cosh^2\bra{\fr{\alpha\mP}{\Lambda^4}}}{1+\tanh\bra{\fr{\alpha\mP}{\Lambda^4}}}-\fr{2\alpha\bar{\mP}}{\Lambda^4}}^{-2}\bar{\na}^\lambda\bar{\mP}\bar{\na}_\lambda\bar{\mP}}.
    \label{conformal_EH_P}
    \ee
Here, $\mP$ on the right-hand side should be regarded as a function of $\bar{\mP}$ through Eq.~\eqref{P2Pbar}.
If we expand the action~\eqref{conformal_EH_P} about $\alpha=0$, we have
    \be
    S[\bar{g}]
    =\int \D^4x\sqrt{-\bar{g}}\,\fr{\Mpl^2}{2}\brb{\bar{R}-\fr{\alpha}{\Lambda^4}\bar{\mP}\bar{R}-\fr{\alpha^2}{\Lambda^8}\bar{\mP}^2\bar{R}+\fr{3\alpha^2}{2\Lambda^8}\bar{\na}^\lambda\bar{\mP}\bar{\na}_\lambda\bar{\mP}+{\cal O}(\alpha^3)},
    \ee
which involves infinitely many higher-derivative interactions.
Therefore, truncated at some finite power of $\alpha$, the theory described by the action~\eqref{conformal_EH_P} would yield Ostrogradsky ghosts.
Nevertheless, due to the invertibility of the conformal transformation, the action~\eqref{conformal_EH_P} as a whole should describe a ghost-free theory.
(See Ref.~\cite{Takahashi:2019vax} for a similar example.)

However, when matter fields are taken into account, the problem of Ostrogradsky ghosts shows up in general.
This is because the matter action does not respect the degeneracy condition that the gravitational action~\eqref{conformal_EH} or \eqref{conformal_EH_P} satisfies.
(See also Refs.~\cite{Deffayet:2020ypa,Takahashi:2021ttd,Takahashi:2022mew,Naruko:2022vuh,Takahashi:2022ctx,Ikeda:2023ntu,Takahashi:2023jro,Takahashi:2023vva} for discussions on a similar problem in the context of scalar-tensor theories.)
Since the ghosts appear only in the presence of matter fields, their mass should scale as some inverse power of the matter energy density.
Therefore, from the EFT viewpoint, the ghosts would be irrelevant at low energies when the energy density of the matter fields is sufficiently small.
In the present paper, we mainly consider a vacuum solution in the theory described by the action~\eqref{conformal_EH_P} and the motion of a test particle around it, and hence the ghosts may be safely neglected.\footnote{However, even in vacuum, quantum effects of matter fields should be take into account in practice.
This could also spoil the degeneracy condition, and we should make sure that the cutoff of the EFT is well above the energy scale of our interest.
Moreover, the correction terms in the effective action could modify a classical solution.
Therefore, for the classical solution to remain valid, the correction terms should have small magnitudes when evaluated for the classical solution.
These requirements would put a bound on the parameter~$\alpha$, and a detailed analysis on this issue is beyond the scope of the present paper.}

%%%%%%%%%%%%%%%%%%%%%%%%%%%%%%%%%%%%%%%%%%%%%%%%%%%%%%%%%%%%%%%%%%%%%%%%%%%%%%%%%%%%
%%%%%%%%%%%%%%%%%%%%%%%%%%%%%%%%%%%%%%%%%%%%%%%%%%%%%%%%%%%%%%%%%%%%%%%%%%%%%%%%%%%%
%	Application to the Kerr metric
%%%%%%%%%%%%%%%%%%%%%%%%%%%%%%%%%%%%%%%%%%%%%%%%%%%%%%%%%%%%%%%%%%%%%%%%%%%%%%%%%%%%
%%%%%%%%%%%%%%%%%%%%%%%%%%%%%%%%%%%%%%%%%%%%%%%%%%%%%%%%%%%%%%%%%%%%%%%%%%%%%%%%%%%%
\subsection{Application to the Kerr metric}\label{sec:conformal_Kerr}

Since the action~\eqref{conformal_EH} is obtained from the Einstein-Hilbert action via the invertible conformal transformation~\eqref{invertible_tr}, any (vacuum) solution in GR is mapped to a solution in the theory described by \eqref{conformal_EH}.
More concretely, for a given solution~$g_\mn=g^{(0)}_\mn$ in GR, we obtain an exact solution in the transformed theory as
    \be
    \bar{g}_\mn=\Omega(\mC^{(0)},\mP^{(0)})g^{(0)}_\mn, \label{conformal_tr_metric}
    \ee
where $\mC^{(0)}$ and $\mP^{(0)}$ denote the quantities~$\mC$ and $\mP$ associated with $g^{(0)}_\mn$, respectively.

\begin{figure}[tb]
  \begin{tabular}{ccc}
  \begin{minipage}{0.26\hsize}
    \begin{center}
    \includegraphics[keepaspectratio=true,height=80mm]{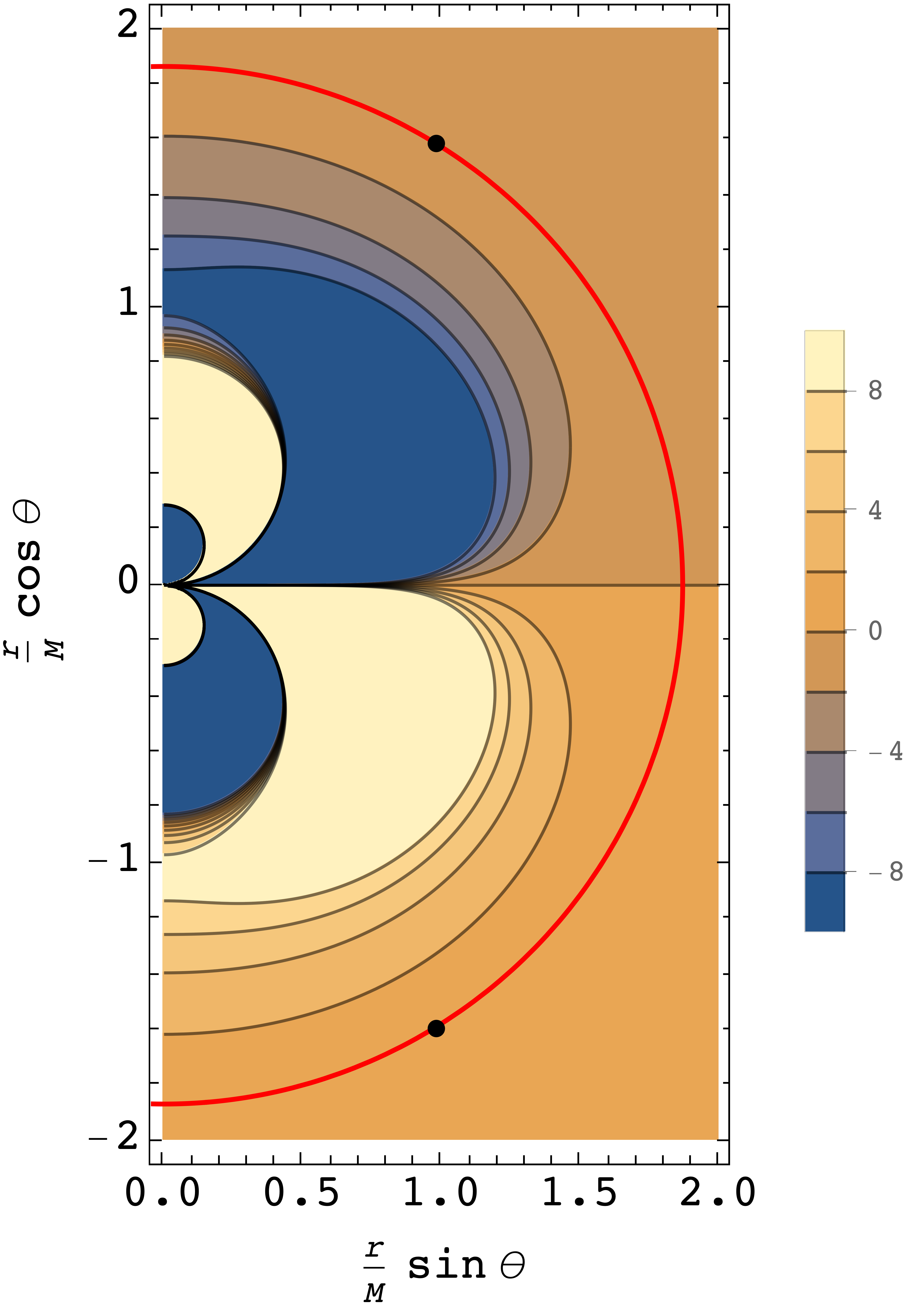}
    \end{center}
  \end{minipage}
  \begin{minipage}{0.23\hsize}
    \begin{center}
    \includegraphics[keepaspectratio=true,height=80mm]{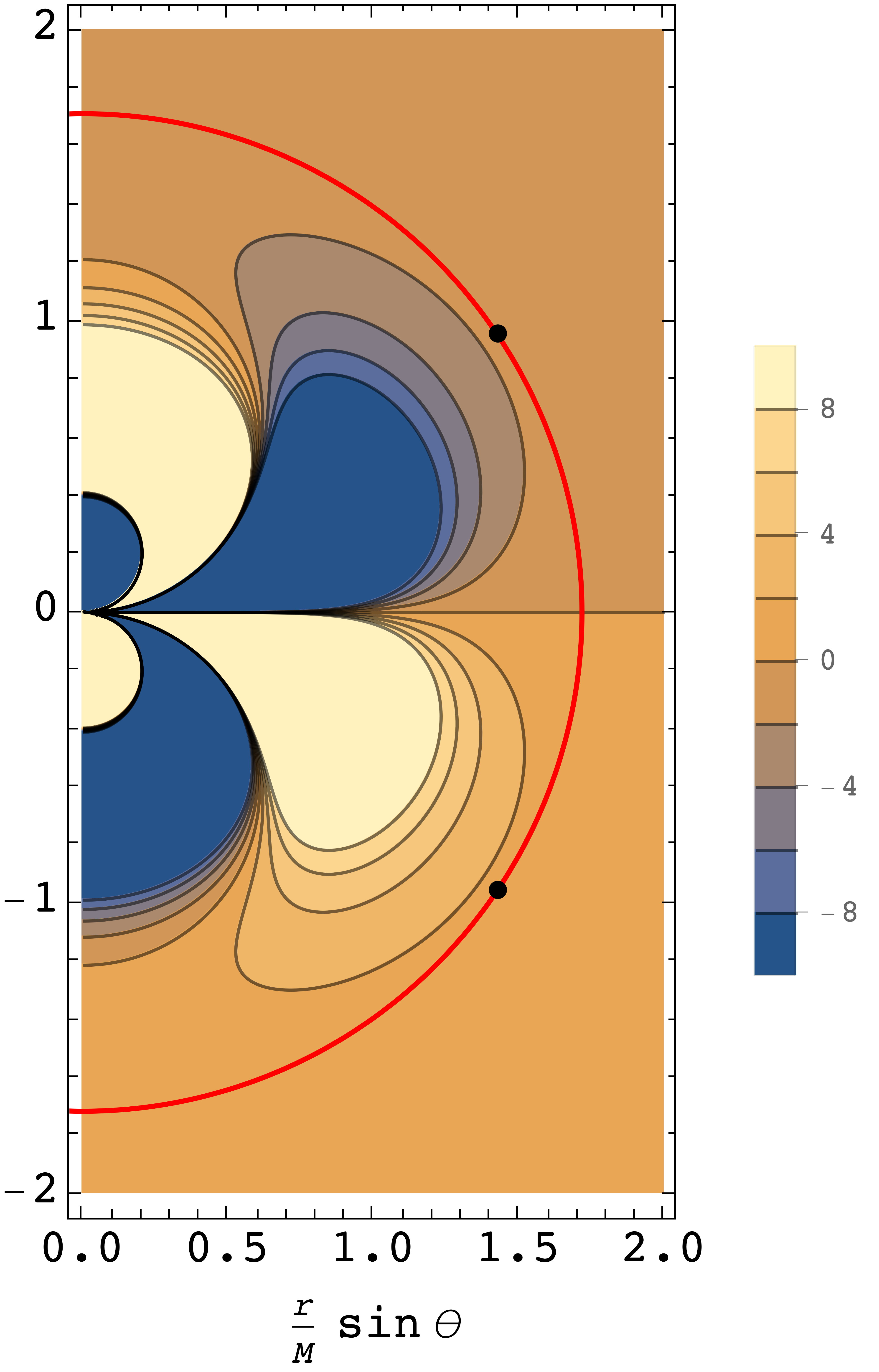}
    \end{center}
  \end{minipage}
  \begin{minipage}{0.23\hsize}
    \begin{center}
    \includegraphics[keepaspectratio=true,height=80mm]{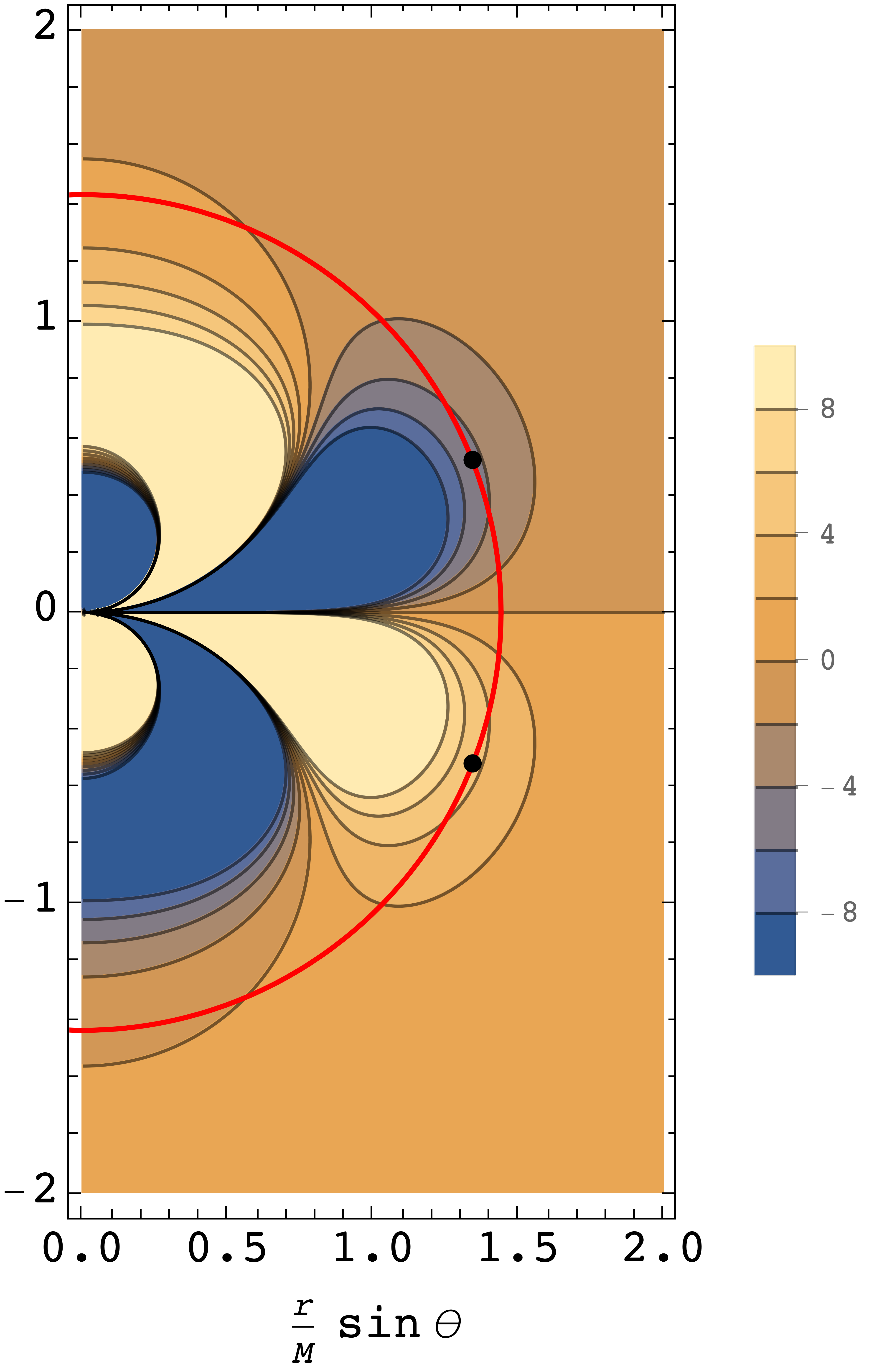}
    \end{center}
  \end{minipage}
  \begin{minipage}{0.25\hsize}
    \begin{center}
    \includegraphics[keepaspectratio=true,height=80mm]{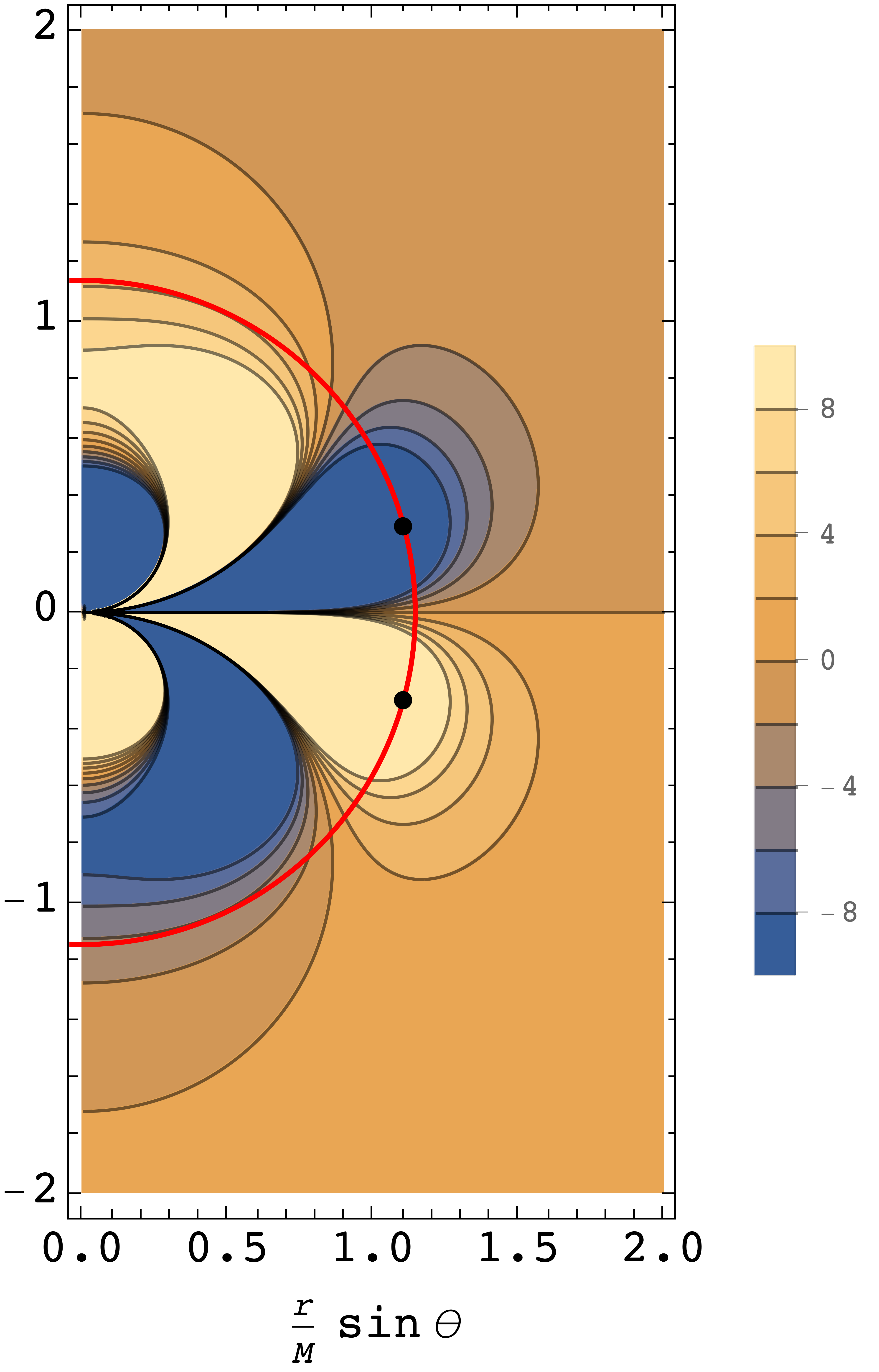}
    \end{center}
  \end{minipage}
  \end{tabular}
  \caption{
       The Chern-Simons term $\mathcal{P}$ given in Eq.~\eqref{eq:CS_term} is plotted for $a=0.5M$, $0.7M$, $0.9M$, and $0.99M$ (from left to right).
       {The color scale is in units of $M^{-4}$.}
       The horizontal and vertical axes represent $r \sin\theta/M$ and $r \cos\theta/M$, respectively. 
       Red circles represent the outer horizon, and black points show the location where $|\mathcal{P}|$ takes its maximum on the horizon.
       }
   \label{fig:CS_xz}
  \end{figure}

In what follows, let us focus on the Kerr metric (i.e., the unique vacuum solution in GR that is stationary, axially symmetric, and asymptotically flat) as the seed of the transformation, {and we shall refer to the transformed metric as the {\it conformal Kerr} metric}.
In terms of the Boyer-Lindquist coordinates~\cite{Carter:1970ea,Carter:1973rla}, the Kerr metric takes the form 
    \be
    {g}_{\mu\nu}\D x^\mu \D x^\nu = 
    - \frac{\Delta}{\rho^2} (\D t-a\sin^2\theta \D\varphi)^2 
    + \frac{\rho^2}{\Delta}\D r^2
    + {\rho^2}\D\theta^2
    + \frac{\sin^2\theta}{\rho^2} \left[ a\D t-(r^2+a^2) \D\varphi \right]^2,
    \ee
where we have defined
    \be
    \rho^2 = r^2+a^2\cos^2\theta, \qquad
    \Delta = r^2+a^2 - 2M r, 
    \ee
with $a$ being the angular momentum per unit mass and $M\,(>0)$ being a parameter of length dimension corresponding to the BH mass.
Note that $\Delta(r)$ has two different real roots~$r_\pm\coloneqq M\pm\sqrt{M^2-a^2}$ for $0<|a|<M$, where $r_+$ and $r_-$ correspond to the radii of outer and inner horizons, respectively.
For the Kerr metric, the quantities~$\mC$ and $\mP$ can be evaluated as
    \be
    \begin{split}
    \mC&=\fr{48M^2(r^6-15a^2r^4\cos^2\theta+15a^4r^2\cos^4\theta-a^6\cos^6\theta)}{\rho^{12}}, \\
    \mP&=-\fr{96aM^2r\cos\theta(r^2-3a^2\cos^2\theta)(3r^2-a^2\cos^2\theta)}{\rho^{12}}.
    \end{split}\label{eq:CS_term}
    \ee
It should be noted that $\mC$ coincides with the Kretschmann scalar because of the Ricci flatness of the Kerr metric,
and 
the Chern-Simons term~$\mP$ is proportional to $a$ and therefore vanishes in the non-rotating limit.
In Fig.~\ref{fig:CS_xz}, we show contour plots of the Chern-Simons term~$\mP$ for different values of $a$.
For reference, we also show the points where $|\mathcal{P}|$ takes its maximum on the horizon (see also Appendix~\ref{appA}).
%-------------------------------------------%
%-------------------------------------------%

Let us now consider the conformal transformation with the conformal factor of the form~\eqref{Omega_P}.
In this case, the conformal Kerr metric
    \be
    \bar{g}_\mn=\brb{1+\tanh\bra{\fr{\alpha\mP}{\Lambda^4}}}g_\mn\qquad
    (\text{$g_\mn$: Kerr metric})
    \label{conformal_Kerr}
    \ee
gives a vacuum solution in the theory described by the action~\eqref{conformal_EH_P}.
Interestingly, the conformal Kerr metric~\eqref{conformal_Kerr} is no longer Ricci flat.
Indeed, the Ricci scalar associated with $\bar{g}_\mn$ is given by
    \begin{align}
    \bar{R}&=-\fr{3\alpha}{\Lambda^4}
    \exp\bra{-\fr{2\alpha\mP}{\Lambda^4}}
    \brc{\Box\mP-\fr{\alpha}{2\Lambda^4}\brb{1+3\tanh\bra{\fr{\alpha\mP}{\Lambda^4}}}\nabla^\alpha\mP\nabla_\alpha\mP} \nonumber \\
    &=\fr{1728\alpha M^2a\cos\theta}{\Lambda^4\rho^{18}}\big[20r\rho^2(16r^6-24r^4\rho^2+10r^2\rho^4-\rho^6) \nonumber \\
    &\qquad\qquad\qquad\qquad~~-M(896r^8-1408r^6\rho^2+640r^4\rho^4-80r^2\rho^6+\rho^8)\big]+\mO(\alpha^2),
    \end{align}
which is nonvanishing unless $a=0$.
This shows that the conformal Kerr metric is not connected to the Kerr metric via a coordinate transformation.
Nevertheless, the conformal Kerr metric remains asymptotically flat.\footnote{In what follows, even when we consider a more general conformal transformation, we always assume that the conformal factor approaches unity for large $r$ so that the metric remains asymptotically flat.}
Also, similarly to the Kerr metric, the curvature singularity shows up at $\rho=0$ only.
We will present expressions of other curvature invariants in Appendix~\ref{appB}.

There are several remarks on the properties of the conformal Kerr solutions obtained in this subsection.
First, the value of $r$ at the outer and inner horizons does not change from that of the original Kerr metric, as the solution is conformally equivalent to the Kerr metric.
Second, the circularity of the spacetime, i.e., the invariance of the metric under $(t,\varphi)\to(-t,-\varphi)$, is maintained.\footnote{On the other hand, the disformal Kerr solution~\cite{Anson:2020trg,BenAchour:2020fgy} does not respect the circularity, which could lead to nontrivial observable effects~\cite{Takamori:2021atp}.}
Finally, the Petrov type of the solution remains unchanged (i.e., remains to be type D) since the Weyl tensor~$\mW^\mu{}_{\nu\lambda\sigma}$ is invariant under a conformal transformation.

%%%%%%%%%%%%%%%%%%%%%%%%%%%%%%%%%%%%%%%%%%%%%%%%%%%%%%%%%%%%%%%%%%%%%%%%%%%%%%%%%%%%
%%%%%%%%%%%%%%%%%%%%%%%%%%%%%%%%%%%%%%%%%%%%%%%%%%%%%%%%%%%%%%%%%%%%%%%%%%%%%%%%%%%%
%	Geodesics
%%%%%%%%%%%%%%%%%%%%%%%%%%%%%%%%%%%%%%%%%%%%%%%%%%%%%%%%%%%%%%%%%%%%%%%%%%%%%%%%%%%%
%%%%%%%%%%%%%%%%%%%%%%%%%%%%%%%%%%%%%%%%%%%%%%%%%%%%%%%%%%%%%%%%%%%%%%%%%%%%%%%%%%%%
\section{Geodesics}\label{sec:geodesics}

In this section, we investigate the geodesic motion of a test particle on the conformal Kerr background.
The trajectory of the particle is denoted by $x^\mu=x^\mu(s)$, where $s$ is an affine parameter and can be identified as the proper time in the case of a timelike particle.
The geodesic equation can be written as
    \be
    \ddot{x}^\lambda + \bar{\Gamma}^\lambda_{\mu\nu} \dot{x}^{\mu} \dot{x}^{\nu}=0, 
    \label{eq:geodesics}
    \ee
where a dot denotes the derivative with respect to $s$ and $\bar{\Gamma}^\lambda_{\mu\nu}$ is the Christoffel symbol associated with the conformal Kerr metric~$\bar{g}_\mn$.
The barred Christoffel symbol is related to the unbarred one (i.e., the one for the Kerr metric) by
    \be
    \bar{\Gamma}^\lambda_{\mu\nu}
    = {\Gamma}^\lambda_{\mu\nu} 
    + \frac{1}{2} \bra{
    \delta^\lambda_\mu \partial_\nu \ln \Omega
    + \delta^\lambda_\nu \partial_\mu \ln \Omega
    - g_{\mu\nu} g^{\lambda\alpha} \partial_\alpha \ln\Omega
    }.
    \ee
Plugging this into Eq.~\eqref{eq:geodesics} yields
    \be
    \Omega(\Omega \dot{x}^\lambda)^{\boldsymbol{\cdot}}
     + \Omega^2  {\Gamma}^\lambda_{\mu\nu} \dot{x}^{\mu} \dot{x}^{\nu}
    = \frac{\kappa}{2}g^{\lambda\alpha} \partial_\alpha \Omega
    \eqqcolon F^\lambda,
    \label{eq:mapped_geodesics} 
    \ee
where we have defined
    \begin{align}
    -\kappa&\coloneqq {\bar g}_{\alpha\beta}\dot{x}^\alpha\dot{x}^\beta
    =\bar{g}_{tt}\dot{t}^2 + 2\bar{g}_{t\varphi}\dot{t}\dot{\varphi} + \bar{g}_{\varphi\varphi}\dot{\varphi}^2 + \bar{g}_{rr} \dot{r}^2 + \bar{g}_{\theta\theta} \dot{\theta}^2,
    \end{align}
and the affine parameter is chosen so that $\kappa=1$ and $0$ for the timelike and null geodesics, respectively.
Interestingly, in terms of a new affine parameter~$\tilde{s}$ such that $\D \tilde{s}=\Omega^{-1}\D s$, we have
    \be
    \fr{\D^2 x^\lambda}{\D\tilde{s}^2}
     +{\Gamma}^\lambda_{\mu\nu} \fr{\D x^{\mu}}{\D\tilde{s}} \fr{\D x^{\nu}}{\D\tilde{s}}
    = F^\lambda,
    \label{eq:mapped_geodesics2}
    \ee
which is nothing but the geodesic equation for the Kerr background with an effective external force~$F^\lambda$.
Note that the nonvanishing components of $F_\lambda$ are only $F_r$ and $F_\theta$ since $\Omega=\Omega(r,\theta)$ in our setup.
Also, $F^\lambda$ vanishes identically for a null geodesic that has $\kappa=0$.

Note that the geodesic equation~\eqref{eq:geodesics} can be equivalently written as
    \be
    (\bar{g}_{\lambda\mu}\dot{x}^\mu)^{\boldsymbol{\cdot}}-\fr{1}{2}\dot{x}^\mu\dot{x}^\nu\pa_\lambda\bar{g}_\mn=0.
    \ee
Together with the fact that the components of the conformal Kerr metric do not depend explicitly on $t$ or $\varphi$, we find that there exist two conserved quantities:
    \be
    \begin{split}
    -E&\coloneqq \bar{g}_{t\mu}\dot{x}^\mu
    =\bar{g}_{tt}\dot{t}+\bar{g}_{t\varphi}\dot{\varphi}, \\
    L_z&\coloneqq \bar{g}_{\varphi\mu}\dot{x}^\mu
    =\bar{g}_{t\varphi}\dot{t}+\bar{g}_{\varphi\varphi}\dot{\varphi},
    \end{split} \label{eq:ELz_def}
    \ee
where $E$ and $L_z$ correspond to the energy and the angular momentum of the particle (per unit mass), respectively.

In what follows, we study the geodesic motion of null or timelike particles in the conformal Kerr spacetime.
Throughout this section, we assume $a>0$ for simplicity.

\subsection{Null geodesics}\label{sec:nullgeodesic}

Let us now consider the case of null geodesics, for which $\kappa=0$.
In this case, the right-hand side of Eq.~\eqref{eq:mapped_geodesics2} vanishes, and hence the geodesic equation has exactly the same form as the one for the Kerr background, reflecting the fact that a null geodesic remains unchanged under a conformal transformation.
As a result, the geodesic equation for a massless particle on the conformal Kerr background is completely integrable, just as in the Kerr case.
Indeed, when $\kappa=0$, there exists a conserved quantity defined by
    \be
    \mQ\coloneqq \bar{g}_{\theta\theta}^2\dot{\theta}^2+\cos^2\theta\bra{\fr{L_z^2}{\sin^2\theta}-a^2E^2},
    \label{calQ}
    \ee
which corresponds to the Carter constant for a massless particle on the Kerr background~\cite{Carter:1968rr}.
Therefore, we obtain the following system of equations:
    \be
    \Omega \rho^2 \dot t
    =a\mW + \fr{(r^2+a^2)\mV}{\Delta_r}, \qquad
    \Omega \rho^2 \dot \varphi
    =\frac{\mW}{\sin^2\theta} + \fr{a \mV}{\Delta_r}, \qquad
    \Omega^2 \rho^4 {\dot r}^2
    =\mV^2 - \Delta_r \mK, \qquad
    \Omega^2 \rho^4 {\dot \theta}^2
    =\mK -\frac{\mW^2}{\sin^2\theta},
    \label{eq:first_integral}
    \ee
where we have defined
    \be
    \mV\coloneqq (r^2+a^2)E - a L_z , \qquad
    \mW\coloneqq L_z - a E \sin^2 \theta, \qquad
    \mK\coloneqq \mQ+(L_z-aE)^2.
    \ee
Since the null geodesic on the conformal Kerr background is the same as the one for the Kerr background, one cannot tell the difference between the two from the motion of a massless test particle.

Note that the conserved quantity~$\mQ$ for a null geodesic is related to a conformal Killing tensor~\cite{Walker:1970un}.
Let $K_\mn$ be the nontrivial Killing tensor~\cite{Walker:1970un} (such that $\nabla_{(\mu}K_{\alpha\beta)}=0$) for the seed Kerr metric, i.e.,
    \be
    K_\mn \D x^\mu \D x^\nu
    =r^2g_\mn\D x^\mu \D x^\nu+\Delta (\D t-a\sin^2\theta\D\varphi)^2-\fr{\rho^4}{\Delta}\D r^2.
    \ee
Then, one can show that $\ti{K}_\mn\coloneqq \Omega^2K_\mn$ is a conformal Killing tensor on the conformal Kerr background.
Indeed, we have
    \be
    \bar{\nabla}_{(\mu}\ti{K}_{\alpha\beta)}
    =\bar{g}_{(\alpha\beta}K^\nu_{\mu)}\pa_\nu\Omega.
    \ee
One can show that the quantity~$\mQ$ defined in Eq.~\eqref{calQ} coincides with $\ti{K}_\mn\dot{x}^\mu\dot{x}^\nu$ for a null geodesic.

\subsection{Timelike circular orbits}\label{sec:timelikegeodesic}

In this subsection, we study timelike circular orbits
[corresponding to $\kappa=1$ in Eq.~\eqref{eq:mapped_geodesics}] that are normal to the symmetry axis, for which $r(s)={\rm const}$ and $\theta(s)={\rm const}$.
As we discuss in Appendix~\ref{appC}, it is useful to define an effective potential by
    \be
    V_{\rm eff}(r,\theta)
    \coloneqq 1+\fr{L_z^2 \bar{g}_{tt}+2EL_z \bar{g}_{t\varphi}+E^2 \bar{g}_{\varphi\varphi}}{\bar{g}_{tt}\bar{g}_{\varphi\varphi}-\bar{g}_{t\varphi}^2}.
    \ee
In terms of this $V_{\rm eff}$, a circular orbit can be realized at $(r,\theta)$ such that $V_{\rm eff}$ and its first derivatives vanish.
From $V_{\rm eff}=0$ and Eq.~\eqref{eq:ELz_def}, we get
    \be
    E=-\fr{\bar{g}_{tt}+\bar{g}_{t\varphi} \omega}{\sqrt{-(\bar{g}_{tt}+2 \bar{g}_{t\varphi} \omega+\bar{g}_{\varphi\varphi} \omega^2)}}, \qquad
    L_z=\fr{\bar{g}_{t\varphi}+\bar{g}_{\varphi\varphi} \omega}{\sqrt{-(\bar{g}_{tt}+2 \bar{g}_{t\varphi} \omega+\bar{g}_{\varphi\varphi} \omega^2)}}.
    \label{eq:ELz}
    \ee
Here, we have defined the orbital angular velocity~$\omega\coloneqq \dot{\varphi}/\dot{t}$, which is constant for a circular orbit.
With the relations~\eqref{eq:ELz}, the functional form of the effective potential~$V_{\rm eff}(r,\theta)$ is determined by $\omega$ only.
Then, $\partial_r V_{\rm eff}=0$ and $\partial_\theta V_{\rm eff}=0$ yield two conditions on $r$, $\theta$, and $\omega$, which can be used to express, e.g., $\theta$ and $\omega$ as functions of $r$.
This means that the expression inside the square root in Eq.~\eqref{eq:ELz} can be regarded as a function of $r$, and it has to be positive so that $E$ and $L_z$ are real.
Actually, there exists a critical radius at which $\bar{g}_{tt}+2 \bar{g}_{t\varphi} \omega+\bar{g}_{\varphi\varphi} \omega^2=0$.
As $r$ approaches this critical radius, the energy~$E$ of the particle diverges, and hence the trajectory of the particle would approach a null geodesic.
It should be noted that a null circular orbit perpendicular to the symmetry axis can be realized only on the equatorial plane, and its radius~$r_{\rm ph}^\pm$ satisfies $r^2-3Mr\pm 2a\sqrt{Mr}=0$ for the prograde~($+$) and retrograde~($-$) orbits, respectively.
This implies that the critical radius mentioned above is nothing but $r=r_{\rm ph}^\pm$.
(Note that the value of $r_{\rm ph}^\pm$ for the conformal Kerr background remains the same as that for the Kerr background as a null geodesic remains unchanged under a conformal transformation.)
For this reason, in what follows, we only consider circular orbits for $r>r_{\rm ph}^\pm$.

Let us discuss the circular orbit in the case where the deviation of $\Omega$ from unity (i.e., the deviation from the Kerr background) is small, without assuming a particular form of $\Omega(r,\theta)$.
In the case of Kerr background which has an equatorial symmetry, one considers circular orbits within the equatorial plane~$\theta=\pi/2$.\footnote{We shall call $\theta=\pi/2$ the equatorial plane, but note that the conformal Kerr spacetime is not symmetric under $\theta\to\pi-\theta$.}
On the other hand, we expect an off-equatorial orbit for our conformal Kerr background.
Indeed, the absence of circular orbits on the equatorial plane~$\theta=\pi/2$ on a parity-violating BH background has been pointed out in Refs.~\cite{Cano:2019ore,Chen:2021ryb}.
However, the deviation from $\theta=\pi/2$ would be small when the deviation from the Kerr background is small.
Assuming $\Omega=1+\ep\varpi(r,\theta)$ and $\theta=\pi/2-\ep\vartheta$ {with $\epsilon$ being a small dimensionless parameter}, the condition~$\partial_r V_{\rm eff}=0$ yields
    \be
    \omega=\omega^\pm
    \coloneqq \fr{1}{a\pm\sqrt{r^3/M}}\brb{1+\ep\fr{r^2(r^2-3Mr\pm 
    2a\sqrt{Mr})}{4M(r^2\pm a\sqrt{Mr})}\pa_r\varpi(r,\pi/2)+\mO(\ep^2)},
    \label{eq:omega}
    \ee
where the positive and negative signs correspond to prograde and retrograde orbits, respectively.
(Note that $a<\sqrt{r^3/M}$ for $r>r_+$, and hence $\omega^-<0$.)
Here and in what follows, the double signs are in the same order.
Then, from $\partial_\theta V_{\rm eff}=0$, we have
    \be
    \ep\vartheta=\ep\vartheta^{\pm}
    \coloneqq \ep\fr{r(r^2-3Mr\pm 2a\sqrt{Mr})}{2M(r^2+3a^2\mp 4a\sqrt{Mr})}\pa_\theta\varpi(r,\pi/2)+\mO(\ep^2).\label{eq:varthetapm}
    \ee
We note that $\pa_\theta\varpi(r,\pi/2)$ sources the deviation of the circular orbit from the equatorial plane~$\theta=\pi/2$, which is expected from the fact that the gradient of $\Omega$ serves as the effective external force [see Eq.~\eqref{eq:mapped_geodesics2}].
Note also that the corrections of $\mO(\ep)$ in Eqs.~\eqref{eq:omega} and \eqref{eq:varthetapm} vanish in the limit of $r\to r_{\rm ph}^\pm$, which is consistent with the fact that the null circular orbit lies on the equatorial plane.

It is also important to study the stability of the orbit.
As clarified in Appendix~\ref{appC}, the information on the stability under linear perturbations is also encoded in the second derivatives of the effective potential~$V_{\rm eff}$.
Written explicitly, the stability conditions read
    \be
    \pa_r^2 V_{\rm eff}\,\pa_\theta^2 V_{\rm eff}-(\pa_r\pa_\theta V_{\rm eff})^2>0 \quad
    \text{and} \quad
    \fr{1}{g_{rr}}\pa_r^2 V_{\rm eff}+\fr{1}{g_{\theta\theta}}\pa_\theta^2 V_{\rm eff}>0.\label{eq:det_and_tr}
    \ee
Note that the left-hand sides of these conditions are regarded as functions of $r$ only, as we have already expressed $\omega$ and $\theta$ in terms of $r$.
As explained earlier, a circular orbit can be realized only for $r>r_{\rm ph}^\pm$, and the orbit would be unstable in the very vicinity of $r=r_{\rm ph}^\pm$.
Also, for large enough $r$ where the conformal factor approaches unity, the spacetime is well described by the Kerr metric and the circular orbit would be stable.
Therefore, we expect that the stability should change at least once (precisely speaking, an odd number of times) within the range~$r_{\rm ph}^\pm<r<\infty$.
In general, if the stability changes at some values of $r\,(>r_{\rm ph}^\pm)$, these radii correspond to marginally stable circular orbits (MSCOs).
In the case of conformal Kerr BHs, as we shall see shortly, there can be multiple MSCOs, and the one with the smallest value of $r$ is the innermost stable circular orbit (ISCO).
When the deviation from the Kerr background is sufficiently small, there exists only a single MSCO, which is also the ISCO.
The deviation from the ISCO radius for the Kerr background, $r_{\rm ISCO}^{(0)}$, is given by\footnote{The ISCO radius for the Kerr background~$r=r_{\rm ISCO}^{(0)\pm}$ is given by the unique positive solution to $r^2-6Mr-3a^2\pm 8a\sqrt{Mr}=0$. For the Schwarzschild case with $a=0$, we have $r_{\rm ISCO}^{(0)+}=r_{\rm ISCO}^{(0)-}=6M$. In the extremal limit~$a\to M$, we have $r_{\rm ISCO}^{(0)+}\to M$ and $r_{\rm ISCO}^{(0)-}\to 9M$.}
    \be
    \fr{r_{\rm ISCO}^\pm-r_{\rm ISCO}^{(0)\pm}}{r_{\rm ISCO}^{(0)\pm}}
    =-\fr{\ep}{4}\bra{1\mp\fr{a}{M\varrho}}^2\brc{\fr{18\varrho(\varrho^2-1)[M^2\varrho^4(\varrho^2-3)+a^2(3\varrho^2-1)]\pa_\varrho\varpi}{M^2\varrho^2(3\varrho^6-4\varrho^4-33\varrho^2+18)+a^2(27\varrho^4-6\varrho^2-5)}+\varrho^2\pa_\varrho^2\varpi}+\mO(\ep^2),
    \label{rISCO}
    \ee
where we have defined $\varrho\coloneqq \sqrt{r/M}$ (so that $\varrho\pa_\varrho=2r\pa_r$ and $\varrho^2\pa_\varrho^2=4r^2\pa_r^2+2r\pa_r$) and the right-hand side is evaluated at $(r,\theta)=(r_{\rm ISCO}^{(0)\pm},\pi/2)$.
In the extremal limit~$a\to M$, we have
    \be
    r_{\rm ISCO}=\left\{
    \begin{array}{ll}
    M\brb{1+\mO(\ep^2)} & \text{ (prograde)}, \\
    9M\brb{1-\ep\bra{6\pa_\varrho\varpi+4\pa_\varrho^2\varpi}+\mO(\ep^2)} & \text{ (retrograde)}.
    \end{array}
    \right.
    \ee

\begin{figure}[tb]
    \begin{center}
        \begin{tabular}{rr}
        \includegraphics[keepaspectratio=true,height=50mm]{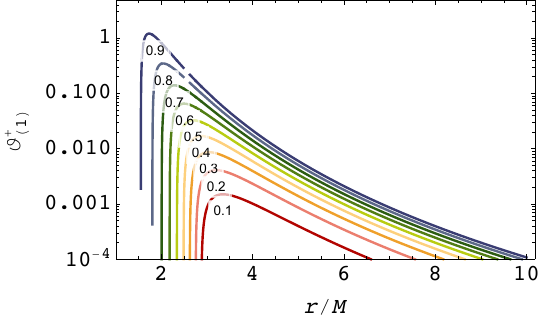}
        \includegraphics[keepaspectratio=true,height=50mm]{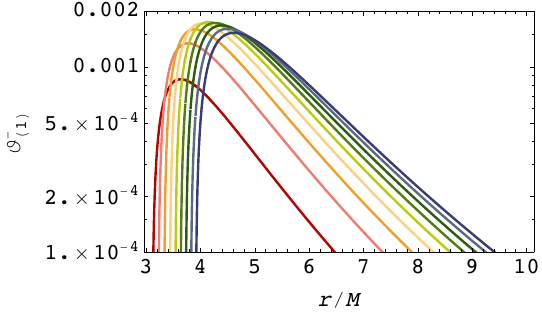}
        \end{tabular}
       \caption{
       The left and right panels respectively show $\vartheta^+_{(1)}$ and $\vartheta^-_{(1)}$ in Eq.~\eqref{eq:vartheta} for different values of $a$, indicated by different colors.
       The numerical value attached on each curve corresponds to the value of $a/M$.
       }
       \label{fig:vartheta}
    \end{center}
  \end{figure}

For concreteness, let us now apply the above analysis to the conformal factor of the form~\eqref{Omega_P}, for which the deviation from the Kerr background is controlled by the dimensionless parameter~$\hat{\alpha}\coloneqq \alpha/(M^4\Lambda^4)$.
In what follows, we assume $\hat{\alpha}>0$ for simplicity.
When $\hat{\alpha}$ is sufficiently small, one can substitute $\hat{\alpha}$ for $\epsilon$ above.
Then, Eq.~\eqref{eq:varthetapm} implies that the value of $\theta$ can be well approximated by $\theta\simeq \pi/2-\hat{\alpha}\vartheta^{\pm}_{(1)}$ with
    \be
    \vartheta^{\pm}_{(1)}\coloneqq
    \fr{144aM^5(r^2-3Mr\pm 
    2a\sqrt{Mr})}{r^6(r^2+3a^2\mp 4a\sqrt{Mr})}.
    \label{eq:vartheta}
    \ee
We plot $\vartheta^{\pm}_{(1)}$ as functions of $r$ in Fig.~\ref{fig:vartheta}.
It should be noted that this approximation can be valid even for $\hat{\alpha}=\mO(1)$ or larger so long as $\hat{\alpha}\vartheta^{\pm}_{(1)}$ is negligible compared to $\pi/2$, which is the case when $r$ is either large enough or close enough to $r_{\rm ph}^\pm$.
Then, the perturbative formulae for $\omega^\pm$ and $r_{\rm ISCO}^\pm$ [Eqs.~\eqref{eq:omega} and \eqref{rISCO}, respectively] would also be valid.
This implies that $\omega^\pm$ and $r_{\rm ISCO}^\pm$ do not receive corrections at the first order in $\hat{\alpha}$ since the $r$-derivatives of $\varpi$ vanish at $\theta=\pi/2$ because of our particular choice of the conformal factor.

%-------------------------------------------%
%-------------------------------------------%

\begin{figure}[htbp]
    \begin{center}
        \begin{minipage}[b]{0.49\textwidth}
        \begin{tabular}{rr}
        \includegraphics[keepaspectratio=true,height=55mm]{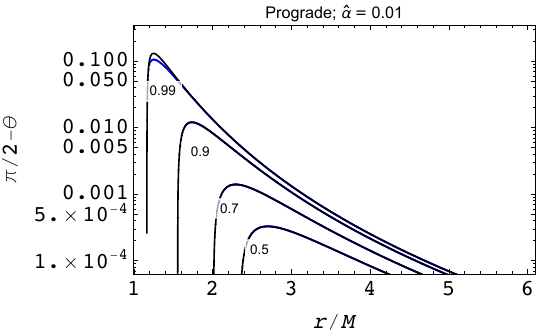}\\
        \includegraphics[keepaspectratio=true,height=55mm]{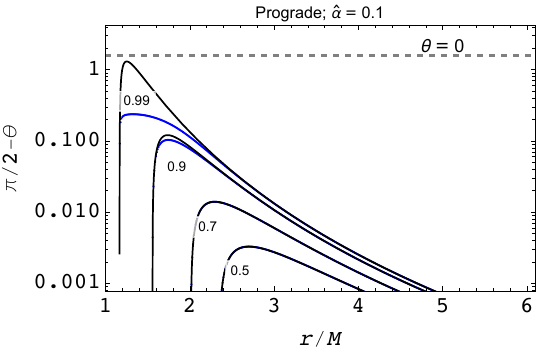}
        \end{tabular}
        \end{minipage}
        \begin{minipage}[b]{0.49\textwidth}
        \begin{tabular}{rr}
        \includegraphics[keepaspectratio=true,height=55mm]{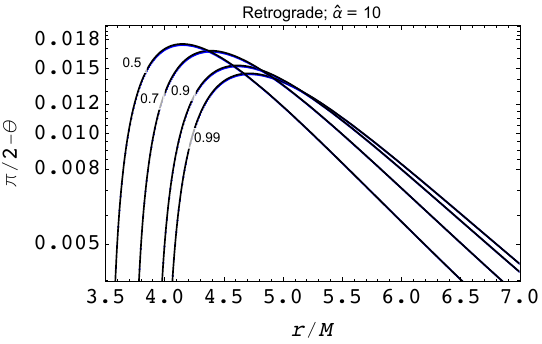}\\
        \includegraphics[keepaspectratio=true,height=55mm]{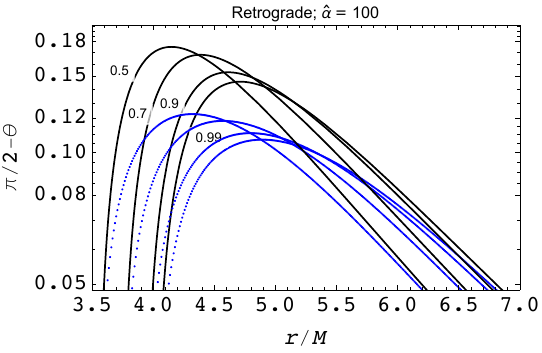}
        \end{tabular}
        \end{minipage}
       \caption{
       The black curves show the value of $\hat{\alpha}\vartheta^\pm_{(1)}$, while the blue curves show the value of $\pi/2-\theta$ obtained numerically.
       The numerical value attached on each curve corresponds to the value of $a/M$.
       }
       \label{fig:numerical_theta}
    \end{center}
  \end{figure}

Figure~\ref{fig:numerical_theta} shows the comparison between $\hat{\alpha}\vartheta^\pm_{(1)}$ and the value of $\pi/2-\theta$ obtained numerically.
Obviously, the approximation~$\theta\simeq \pi/2-\hat{\alpha}\vartheta^{\pm}_{(1)}$ breaks down when $|\hat{\alpha}\vartheta^\pm_{(1)}|$ becomes comparable to $\pi/2$.
The figure shows that the approximation works well up to $\hat{\alpha}\approx 0.01$ in the prograde case, while it works up to $\hat{\alpha}\approx 10$ in the retrograde case.

%-------------------------------------------%
%-------------------------------------------%

\begin{figure}[tb]
  \begin{center}
      \begin{tabular}{cc}
      \includegraphics[keepaspectratio=true,height=50mm]{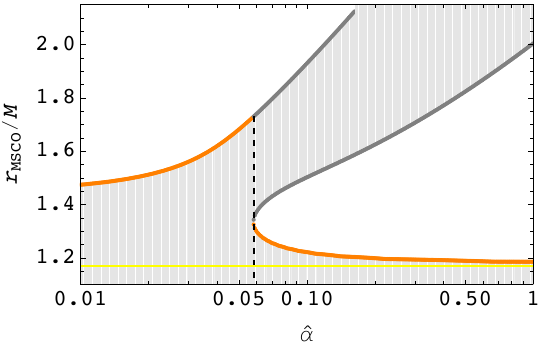}
      \includegraphics[keepaspectratio=true,height=50mm]{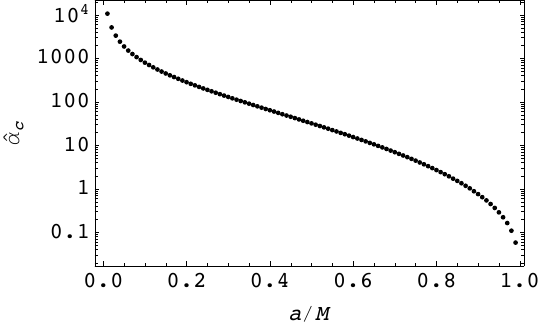}
      \end{tabular}
      \caption{
      The left panel shows the MSCO radii~$r_{\rm MSCO}$ for the prograde orbit around the conformal Kerr BH with $a=0.99M$.
      For the values of $\hat{\alpha}$ larger than the critical value~$\hat{\alpha}_{\rm c}\approx 0.058$ (indicated by the vertical dashed line), there exist three MSCOs, and stable circular orbits do not exist in the shaded region.
      The ISCO radius is indicated by an orange piecewise curve.
      The yellow horizontal line corresponds to $r=r_{\rm ph}^+$.
      In the right panel, the critical value~$\hat{\alpha}_{\rm c}$ is plotted as a function of $a$.
      }
      \label{fig:MSCO}
  \end{center}
\end{figure}

One can also study the stability of the circular orbit based on the conditions~\eqref{eq:det_and_tr}.
For small values of $\hat{\alpha}$, there exists only one MSCO, which is nothing but the ISCO.
Interestingly, there exists a critical value of $\hat{\alpha}$ (called $\hat{\alpha}_{\rm c}$) above which three MSCOs appear, as shown in Fig.~\ref{fig:MSCO}.
The left panel shows the value(s) of $r_{\rm MSCO}$ for the prograde orbit around the conformal Kerr BH with $a=0.99M$ for different values of $\hat{\alpha}$, where the critical value is given by $\hat{\alpha}_{\rm c}\approx 0.058$.
The ISCO for each $\hat{\alpha}$ is indicated by an orange piecewise curve.
This means that, for $\hat{\alpha}>\hat{\alpha}_{\rm c}$, there is an interval of $r$ outside the ISCO where stable circular orbits do not exist, which is a crucial difference from the Kerr case.
In general, the critical value~$\hat{\alpha}_{\rm c}$ is a monotonically decreasing function of $a$, as shown in the right panel of Fig.~\ref{fig:MSCO}.
For the retrograde orbit, such a critical value of $\hat{\alpha}$ also exists, which is much larger (by several orders of magnitude) than that for the prograde orbit for a fixed value of $a$.

%-------------------------------------------%
%-------------------------------------------%

\subsection{Numerical experiments}\label{sec:numerical}

Let us present several numerical experiments for the geodesic motion of a timelike test particle.
Throughout this subsection, we focus on the conformal factor of the form~\eqref{Omega_P}.
Also, for demonstration purposes, we only consider the case with $a>0$ and $\hat{\alpha}>0$.

%-------------------------------------------%
\begin{figure}[tb]
  \begin{center}
      \includegraphics[keepaspectratio=true,height=50mm]{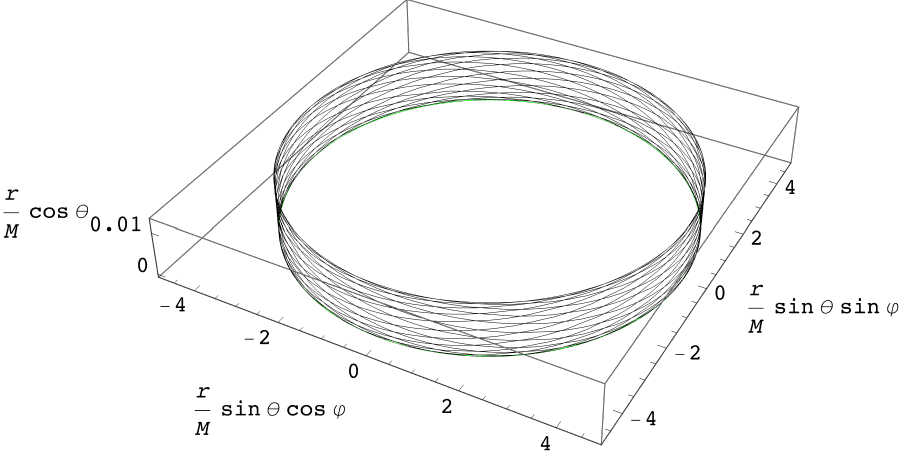}
      \includegraphics[keepaspectratio=true,height=50mm]{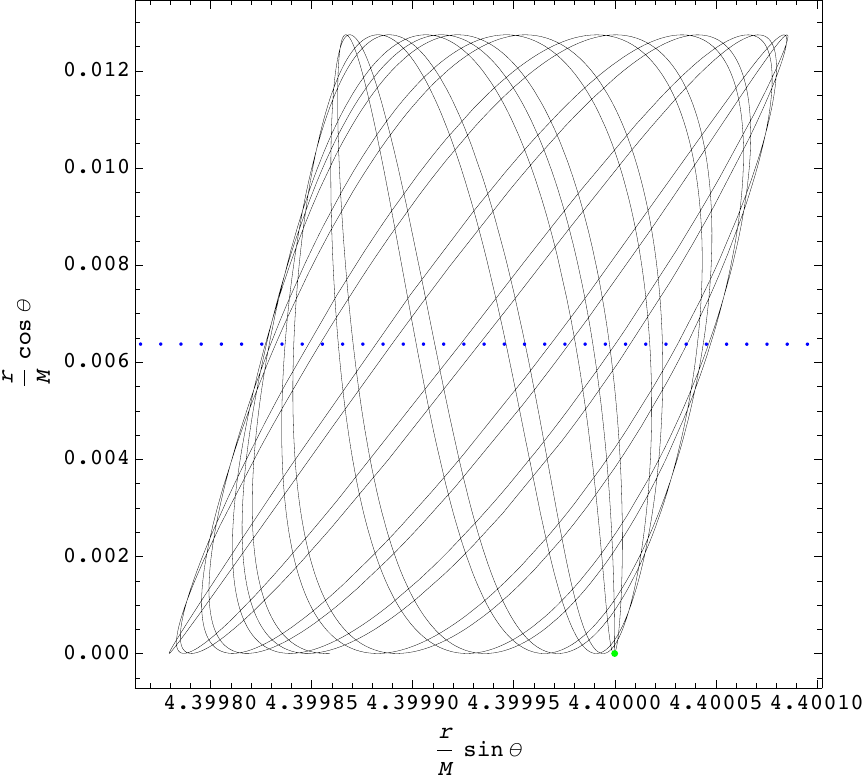}
      \caption{ 
      Orbits of a timelike test particle initially on the equatorial plane in the birds-eye view (left) and projected onto the plane corotating with the particle (right).
      The black curve corresponds to the particle orbit around the conformal Kerr BH with $a=0.99M$ and $\hat{\alpha}=0.1$, where the initial condition is chosen so that the particle would be in an equatorial prograde circular orbit with $r=4.4M$ in the case of Kerr background in GR.
      In the right panel, the particle is oscillating around the blue dotted curve on which circular orbits can be realized.
      For reference, the particle motion for $\hat{\alpha}=0$ (Kerr in GR) is also shown by the green curve and it stays in the equatorial circular orbit.
      }
  \label{fig:perturbedCircularOrbit_Plot3D}
  \end{center}
\end{figure}
%-------------------------------------------%

Figure~\ref{fig:perturbedCircularOrbit_Plot3D} shows the motion of a timelike test particle initially on the equatorial plane around the conformal Kerr BH with $a=0.99M$ and $\hat{\alpha}=0.1$.
The initial condition is chosen so that the particle is in an equatorial prograde circular orbit with $r=4.4M$ in the case of Kerr background with the same value of $a$.
The left panel shows that the particle orbit around the conformal Kerr BH (black) deviates from the equatorial plane, while the one around the Kerr BH (green) remains on the equatorial plane.
The right panel shows the particle orbit projected onto the plane corotating with the particle, which is useful to keep track of the distance from the symmetry axis.
If the particle is in a circular motion, it stays at a point on the corotating plane.
Hence, the Kerr case is represented by the green point located at $(r \sin\theta, r \cos\theta) = (4.4M,0)$.
In the conformal Kerr case, as mentioned earlier, the particle experiences the effective external force~$F_\mu\propto \pa_\mu\Omega$ and does not stay on the equatorial plane.
However, as can be seen in the right panel, the particle oscillates around some particular point on the corotating plane and is almost in a circular motion with $r\approx 4.4M$.
Actually, the center of the oscillation corresponds to the value of $(r,\theta)$ that realizes the circular motion on the conformal Kerr background.
For reference, we plot $(r\sin\theta,r\cos\theta)$ for circular orbits with the blue dotted curve in the right panel.
Note that $(\hat{\alpha},r)=(0.1,4.4M)$ lies within the stable region in the left panel of Fig.~\ref{fig:MSCO}, and hence it is reasonable that the particle oscillates around this circular orbit.

\begin{figure}[tb]
  \begin{center}
      \begin{minipage}[t]{0.49\textwidth}
      \begin{tabular}{cc}
      \includegraphics[keepaspectratio=true,height=90mm]{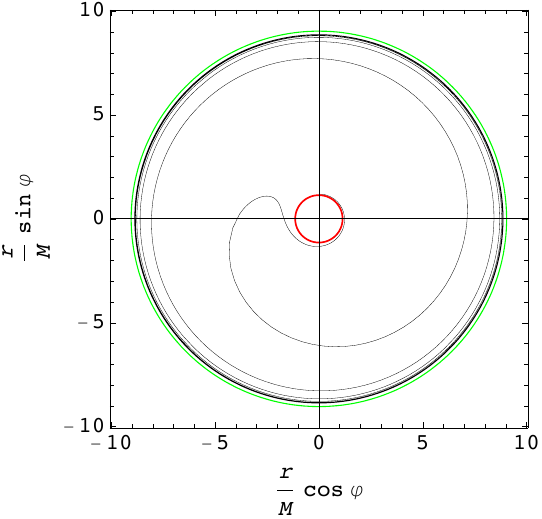}\\\\
      \includegraphics[keepaspectratio=true,height=50mm]{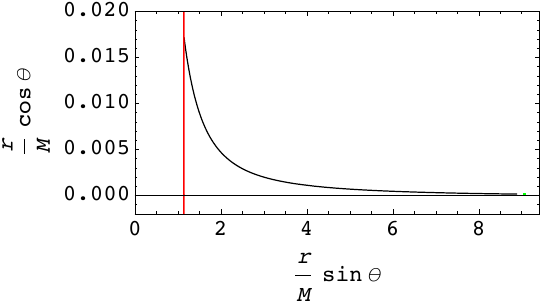}
      \end{tabular}
      \end{minipage}
      \caption{
      Orbits of a timelike test particle initially in a retrograde circular motion around the conformal Kerr BH with $a=0.99M$ and $\hat{\alpha}=0.1$.
      The top panel is the projection of the orbits onto the equatorial plane~$\theta=\pi/2$ and the bottom panel is the projection onto the plane corotating with the particle.
      The green and black curves correspond to the particle orbits with $r_{\rm ini}=(1\pm 10^{-2})r_{\rm ISCO}$, respectively.
      The initial condition for $\dot{\varphi}$ is chosen to be $10^{-9}$ smaller than the value for an exactly circular motion, so that the particle plunges into the horizon (red circle) for an unstable circular orbit.
      }
      \label{fig:ISCO_xy_retrograde}
  \end{center}
\end{figure}

In Fig.~\ref{fig:ISCO_xy_retrograde}, we plot the motion of a test particle with the initial condition chosen so that the particle would be in a retrograde (almost) circular motion perpendicular to the symmetry axis.
Note that the initial condition for $\dot{\varphi}$ is slightly detuned from the value for an exactly circular motion.
Specifically, we have chosen $\dot{\varphi}_{\rm ini} =(1-10^{-9})\dot{\varphi}_{\rm ini,c}$, where $\dot{\varphi}_{\rm ini,c}$ is the value for an exactly circular motion, so that the particle would plunge into the BH if the circular orbit is unstable.
The background spacetime is the same as in Fig.~\ref{fig:perturbedCircularOrbit_Plot3D} (i.e., the conformal Kerr BH with $a=0.99M$ and $\hat{\alpha}=0.1$).
In this case, the retrograde orbits have only one MSCO, which is the ISCO with $r_{\rm ISCO}\approx 8.97M$.
We consider the situation where the particles are initially put near the ISCO radius, $r_{\rm ini}=(1\pm 10^{-2})r_{\rm ISCO}$.
The top panel clearly shows that the particle is in an almost circular orbit if $r_{\rm ini}>r_{\rm ISCO}$ (green curve), while it plunges into the BH if $r_{\rm ini}<r_{\rm ISCO}$ (black curve).
As the plunging particle approaches the BH horizon, the angular momentum of the particle is aligned with the BH spin and the deviation from the equatorial plane becomes sizable.
%-------------------------------------------%
%-------------------------------------------%

\begin{figure}[tb]
  \begin{center}
      \begin{minipage}[t]{0.49\textwidth}
      \begin{tabular}{rr}
      \includegraphics[keepaspectratio=true,height=80mm]{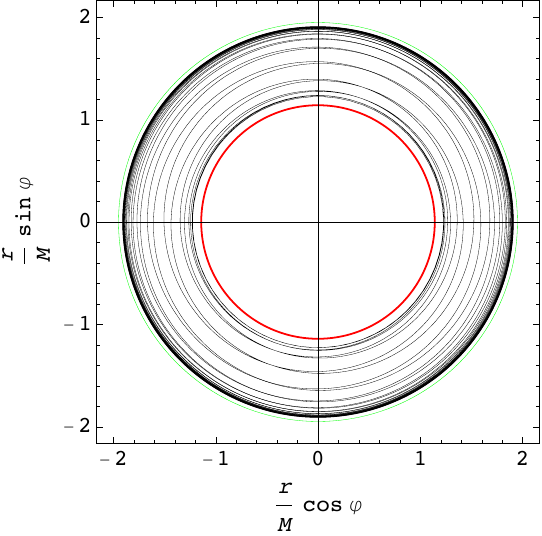}\\\\
      \includegraphics[keepaspectratio=true,height=45mm]{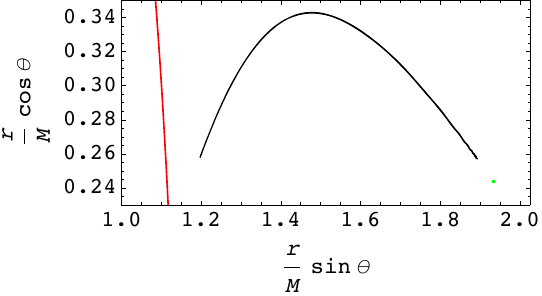}
      \end{tabular}
      \end{minipage}
      \begin{minipage}[t]{0.49\textwidth}
      \begin{tabular}{rr}
      \includegraphics[keepaspectratio=true,height=80mm]{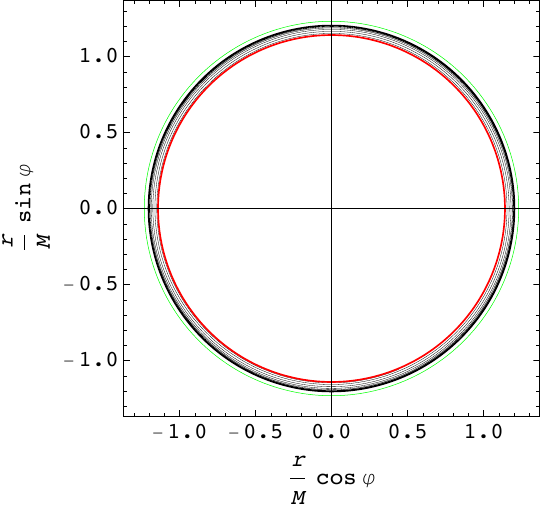}\\\\
      \includegraphics[keepaspectratio=true,height=45mm]{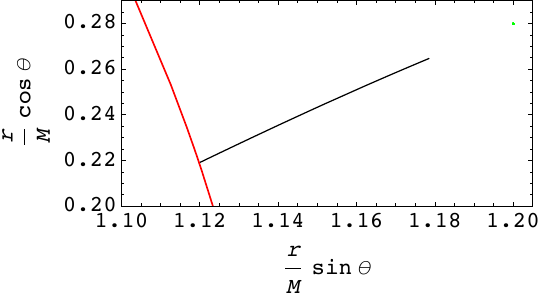}
      \end{tabular}
      \end{minipage}
      \caption{
      Orbits of a timelike test particle initially in a prograde circular motion around the conformal Kerr BH with $a=0.99M$ and $\hat{\alpha}=0.1$.
      The top row is the projection of the orbits onto the equatorial plane~$\theta=\pi/2$ and the bottom row is the projection onto the plane corotating with the particle.
      The left column shows the plots for $r_{\rm ini}\approx 1.93M$ (outermost MSCO) and the right column shows the plots for $r_{\rm ini}\approx 1.22M$ (innermost MSCO, or ISCO).
      The green and black curves correspond to the particle orbits with $r_{\rm ini}=(1\pm 10^{-2})r_{\rm MSCO}$, respectively.
      The initial condition for $\dot{\varphi}$ is chosen to be $10^{-9}$ smaller than the value for an exactly circular motion, so that the particle starts to plunge into the horizon (red circle) for an unstable circular orbit.
      In the left column, the particle oscillates within $1.2M\lesssim r\lesssim 1.9M$ and does not fall into the horizon.
      On the other hand, in the right column, the particle falls into the horizon.
      }
      \label{fig:ISCO_xy_prograde}
  \end{center}
\end{figure}

Similar plots for prograde orbits are shown in Fig.~\ref{fig:ISCO_xy_prograde}.
In this case, as mentioned earlier, there are three MSCO radii~$r_{\rm MSCO}\approx 1.22M$, $1.50M$, and $1.93M$, of which the innermost one corresponds to the ISCO.
Hence, stable circular orbits can be realized for $1.22M\lesssim r\lesssim 1.50M$ and $r\gtrsim 1.93M$ (see the left panel of Fig.~\ref{fig:MSCO}).
The plots in the left column are for $r_{\rm ini}=(1\pm 10^{-2})\times 1.93M$, while those in the right column are for $r_{\rm ini}=(1\pm 10^{-2})\times 1.22M$.
In either case, the particle is in an almost circular orbit if $r_{\rm ini}>r_{\rm MSCO}$, while it is not if $r_{\rm ini}<r_{\rm MSCO}$.\footnote{The opposite situation occurs for $r_{\rm MSCO}\approx 1.50M$ (i.e., the intermediate MSCO radius).
Namely, the orbit is stable for $r_{\rm ini}<r_{\rm MSCO}$ and unstable for $r_{\rm ini}>r_{\rm MSCO}$.}
Interestingly, if the particle is initially put slightly inside $r\approx 1.93M$ (i.e., the outermost MSCO radius), the particle oscillates within $1.2M\lesssim r\lesssim 1.9M$ and does not fall into the BH horizon as shown by black curves in the left column in Fig.~\ref{fig:ISCO_xy_prograde}.
%-------------------------------------------%
%-------------------------------------------%

\begin{figure}[tb]
  \begin{center}
      \begin{minipage}[b]{0.49\textwidth}
          \includegraphics[keepaspectratio=true,height=84mm]{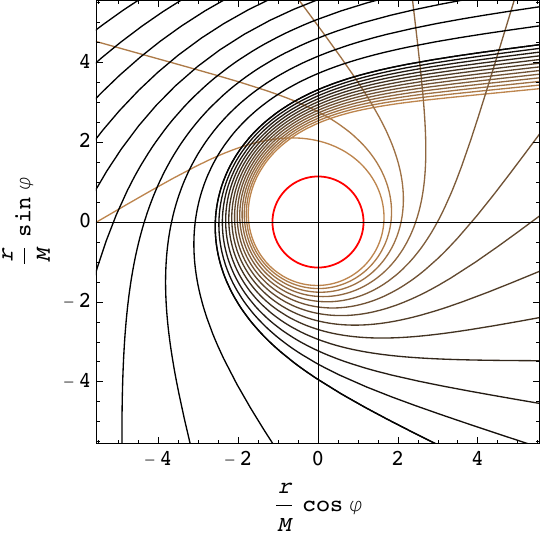}
      \end{minipage}
      \begin{minipage}[b]{0.49\textwidth}
          \includegraphics[keepaspectratio=true,height=42mm]{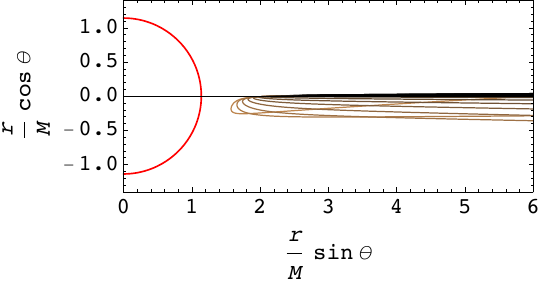}
          \includegraphics[keepaspectratio=true,height=42mm]{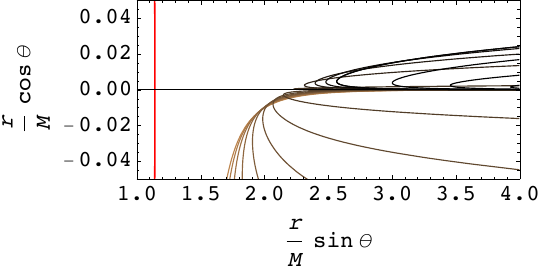}
      \end{minipage}
      
      \caption{
      Scattering of timelike test particles by the conformal Kerr BH with $a=0.99M$ and $\hat{\alpha}=1$.
      The left panel is the projection of the orbits onto the equatorial plane~$\theta=\pi/2$ and the right panels are the projection onto the plane corotating with the particle.
      The right bottom panel is an enlarged view of the top panel.
      The red circle represents the BH horizon.
      Each particle starts on the equatorial plane with the same initial velocity but with different impact parameters.
      The initial angular momentum is aligned with the BH spin.
      }
      \label{fig:sanran_xy}
  \end{center}
\end{figure}

Figure~\ref{fig:sanran_xy} shows the the motion of timelike test particles scattered by the conformal Kerr BH with $a=0.99M$ and $\hat{\alpha}=1$, where we expect that the motion of the particles would be significantly modified from the Kerr case.
Each particle starts on the equatorial plane with the same initial velocity but with different impact parameters.
Similarly to Fig.~\ref{fig:perturbedCircularOrbit_Plot3D}, the particles do not stay on the equatorial plane, and the scattering polar angle depends on the impact parameter.
As it should be, the smaller the impact parameter is, the larger the effects of parity violation are.
%-------------------------------------------%
%-------------------------------------------%

%%%%%%%%%%%%%%%%%%%%%%%%%%%%%%%%%%%%%%%%%%%%%%%%%%%%%%%%%%%%%%%%%%%%%%%%%%%%%%%%%%%%
%%%%%%%%%%%%%%%%%%%%%%%%%%%%%%%%%%%%%%%%%%%%%%%%%%%%%%%%%%%%%%%%%%%%%%%%%%%%%%%%%%%%
%	Conclusions
%%%%%%%%%%%%%%%%%%%%%%%%%%%%%%%%%%%%%%%%%%%%%%%%%%%%%%%%%%%%%%%%%%%%%%%%%%%%%%%%%%%%
%%%%%%%%%%%%%%%%%%%%%%%%%%%%%%%%%%%%%%%%%%%%%%%%%%%%%%%%%%%%%%%%%%%%%%%%%%%%%%%%%%%%
\section{Conclusions}\label{sec:conc}

Recently, there have been growing interests in the possibility of parity violation in the history of the Universe as well as in the vicinity of astrophysical objects.
In Ref.~\cite{Takahashi:2022mew}, ghost-free pure metric theories of gravity with higher-order derivatives have been obtained by performing an invertible conformal transformation on the Einstein-Hilbert action, which we have reviewed in \S\ref{sec:trs}.
When the conformal transformation depends on the Chern-Simons term~$\mP$, the resultant theory involves parity-violating interactions in general.
Importantly, under this transformation, any vacuum solution in GR is mapped to that in the transformed theory.
We have chosen the Kerr solution in GR as a seed and obtained a novel class of {\it conformal Kerr} metrics as an exact solution in the parity-violating gravitational theories.
We emphasize that our construction does not rely on either slow-rotation or near-extremal approximations.

In \S\ref{sec:geodesics}, we have investigated the geodesic motion of a test particle in the conformal Kerr spacetime.
For a null geodesic, as discussed in \S\ref{sec:nullgeodesic}, the geodesic equation remains the same as that in the Kerr spacetime.
Therefore, one cannot tell the difference between these two spacetimes from the motion of photon.
On the other hand, for a timelike geodesic, the geodesic equation receives a nontrivial correction, which can be understood as an effective external force.
The effective external force can have a nonvanishing $\theta$-component, and hence, e.g., circular orbits perpendicular to the symmetry axis do not lie in the equatorial plane.
In \S\ref{sec:timelikegeodesic}, we have derived a perturbative formula for the deviation of such circular orbits from the equatorial plane.
Interestingly, the deviation from the equatorial plane is different for the prograde and retrograde circular orbits, which would offer a possibility to detect the equatorial asymmetry of the conformal Kerr spacetime, e.g., through observations of orbiting stars. 
We have also studied the stability of the circular orbits and found that the conformal Kerr BHs can have multiple MSCOs, which is a crucial difference from the Kerr case.
Finally, in \S\ref{sec:numerical}, we have presented several numerical experiments that highlight the difference from the case of Kerr background in GR.

There are several possible future studies.
While the null geodesic in the conformal Kerr spacetime is the same as that in the Kerr spacetime, it does not necessarily mean that observed BH shadow remains unchanged since the accretion of matter fields should be taken into account in reality.
It would be intriguing to reveal how effects of the deviation from the Kerr spacetime are imprinted in BH shadow.
It would also be interesting to investigate perturbations of the conformal Kerr solution in the parity-violating theory.
In particular, the study of quasinormal modes as well as the propagation of gravitational waves in the conformal Kerr spacetime would provide more insight into the possible parity violation in our Universe.
These issues are left for future work.

%%%%%%%%%%%%%%%%%%%%%%%%%%%%%%%%%%%%%%%%%%%%%%%%%%%%%%%%%%%%%%%%%%%%%%%%%%%%%%%%%%%%
%%%%%%%%%%%%%%%%%%%%%%%%%%%%%%%%%%%%%%%%%%%%%%%%%%%%%%%%%%%%%%%%%%%%%%%%%%%%%%%%%%%%

\acknowledgments{
We would like to thank Che-Yu Chen, Keisuke Nakashi, and Kazumasa Okabayashi for useful discussions.
This work was supported by JSPS (Japan Society for the Promotion of Science) KAKENHI Grant Nos.~JP21F21019 (H.W.H.T.), JP22KJ1646 (K.T.), JP23K13101 (K.T.), and JP22K03639 (H.M.).
M.M.~was supported by the Portuguese national fund through the Funda\c{c}\~{a}o para a Ci\^encia e a Tecnologia in the scope of the framework of the Decree-Law 57/2016 of August 29, changed by Law 57/2017 of July 19, and the Centro de Astrof\'{\i}sica e Gravita\c c\~ao through the Project~No.~UIDB/00099/2020.}

%%%%%%%%%%%%%%%%%%%%%%%%%%%%%%%%%%%%%%%%%%%%%%%%%%%%%%%%%%%%%%%%%%%%%%%%%%%%%%%%%%%%
%%%%%%%%%%%%%%%%%%%%%%%%%%%%%%%%%%%%%%%%%%%%%%%%%%%%%%%%%%%%%%%%%%%%%%%%%%%%%%%%%%%%

\appendix

\section{Chern-Simons scalar for the Kerr metric}
\label{appA}

In \S\ref{sec:conformal_Kerr}, we studied a conformal transformation of the Kerr metric, where the conformal factor is a function of the Chern-Simons term~$\mP$ [see Eq.~\eqref{conformal_Kerr}].
As discussed in \S\ref{sec:trnsf_law}, the transformation is invertible so long as we are interested in the region where $\mP$ is finite (i.e., unless we consider a curvature singularity).
However, when such a transformation is applied to the Kerr metric (or any other BH metrics), it would suffice if the invertibility condition is satisfied (on and) outside the horizon:
Namely, the invertibility condition may be violated inside the horizon.
In this case, in order to see whether the invertibility condition is satisfied, we have to know the possible range of $\mP$ outside the horizon~$r=r_+\coloneqq M+\sqrt{M^2-a^2}$.
Therefore, in this appendix, we present an analytic expression for the maximal value of $|\mP|$.\footnote{Regarding the Weyl tensor squared~$\mC$ defined in \eqref{CP}, its absolute value is maximized at $(r,\cos\theta)=(r_+,\pm 1)$ and the maximal value is given by $|\mC|_{\rm max}=48M^2/r_+^6$.}

Let us introduce a dimensionless parameter~$\delta\coloneqq (r_{+}-M)/M=\sqrt{M^2-a^2}/M$.
It should be noted that $0<\delta<1$ for $0<|a|<M$ and $\delta\to 0$ corresponds to the extremal limit~$|a|\to M$.
Let $\delta_0$ be the largest solution to $8\delta^3-4\delta^2-4\delta+1=0$, i.e.,
    \be
    \delta_0
    \coloneqq \fr{1}{6}+\fr{\sqrt{7}}{3}\cos\bra{\fr{1}{3}\cos^{-1}\fr{-1}{2\sqrt{7}}}
    \approx 0.900969.
    \ee
Note that $\delta=\delta_0$ corresponds to $|a|\approx 0.433884M$.
Then, the maximal value of $|\mP|$ in the region~$r\ge r_+$ and $0\le \theta\le \pi$ is given by
    \be
    |\mP|_{\rm max}=\left\{
    \begin{array}{lll}
    \displaystyle\fr{6(4\delta_0^2-1)(1+\delta_0)^4}{M^4(1+\delta)^6}\sqrt{\fr{1-\delta_0}{1+\delta_0}}
    &\displaystyle\text{ at }(r,\cos\theta)=\bra{r_+,\pm\sqrt{\fr{(1-\delta_0)(1+\delta)}{(1+\delta_0)(1-\delta)}}}
    &\text{ for }0<\delta\le\delta_0, \\
    \displaystyle\fr{6(4\delta^2-1)}{M^4(1+\delta)^2}\sqrt{\fr{1-\delta}{1+\delta}}
    &\displaystyle\text{ at }(r,\cos\theta)=\bra{r_+,\pm 1}
    &\text{ for }\delta_0\le\delta<1.
    \end{array}
    \right.
    \ee
Here, $|\mP|_{\rm max}$ is a monotonically decreasing function of $\delta$ (i.e., a monotonically increasing function of $|a|$).
In the extremal limit~$\delta\to 0$ or $|a|\to M$, the maximal value of $|\mP|$ is given by
    \be
    |\mP|_{\rm max}=\fr{49}{8M^4}\brb{\sqrt{7}+4\cos\bra{\fr{1}{3}\cos^{-1}\fr{211}{98\sqrt{7}}}}
    \approx 40.1836 M^{-4}
    \qquad (\text{extremal limit $|a|\to M$}).
    \ee

%%%%%%%%%%%%%%%%%%%%%%%%%%%%%%%%%%%%%%%%%%
\section{Curvature invariants of the conformally transformed metric}
\label{appB}

In this appendix, we present the expression for curvature invariants of the conformally transformed metric 
\be \bar g_{\mu\nu} = \Omega g_{\mu\nu} , \ee
with $\Omega$ being an arbitrary nonvanishing function of spacetime.
In general, we have the following formulae for the transformation law of curvature invariants:
    \begin{align}
    \bar{R}^{\alpha\beta\gamma\delta}\bar{R}_{\alpha\beta\gamma\delta}
    &=\fr{1}{\Omega^2}\brb{\mC -R^\alpha_\beta\bra{4\fr{\Omega^\beta_\alpha}{\Omega}-6\fr{\Omega_\alpha\Omega^\beta}{\Omega^2}}
    -\fr{R \Omega_{\alpha}\Omega^{\alpha}}{\Omega^2}
    +2\fr{\Omega^\alpha_\beta\Omega^\beta_\alpha}{\Omega^2}-6\fr{\Omega^\alpha\Omega^\beta\Omega_{\alpha\beta}}{\Omega^3}+\bra{\fr{\Box\Omega}{\Omega}}^2+\fr{15}{4}\bra{\fr{\Omega^\alpha\Omega_\alpha}{\Omega^2}}^2}, \nonumber \\
    \bar{R}^{\alpha\beta}\bar{R}_{\alpha\beta}
    &=\fr{1}{\Omega^2}\left[\sizecorr{\bra{\fr{\Box\Omega}{\Omega}}^2}
    R^\alpha_\beta\bra{R^\beta_\alpha+ 3\fr{\Omega_{\alpha}\Omega^{\beta}}{\Omega^2}-2\fr{\Omega_{\alpha}^{\beta}}{\Omega}}-\fr{R\Box\Omega}{\Omega}
    +\fr{\Omega^\alpha_\beta\Omega^\beta_\alpha}{\Omega^2}-3\fr{\Omega^\alpha\Omega^\beta\Omega_{\alpha\beta}}{\Omega^3}+2\bra{\fr{\Box\Omega}{\Omega}}^2-\fr{3}{2}\fr{\Omega^\alpha\Omega_\alpha\Box\Omega}{\Omega^3}\right. \nonumber \\
    &\qquad\qquad\left.+\fr{9}{4}\bra{\fr{\Omega^\alpha\Omega_\alpha}{\Omega^2}}^2\right], \nonumber \\
    \bar{R}
    &=\fr{1}{\Omega}\bra{R-3\fr{\Box\Omega}{\Omega}+\fr{3}{2}\fr{\Omega^\alpha\Omega_\alpha}{\Omega^2}},
    \label{curv_inv_conformal}
    \end{align}
where $\Omega_\alpha\coloneqq\na_\alpha\Omega$ and $\Omega_{\alpha\beta}\coloneqq\na_\alpha\na_\beta\Omega$.
As a consistency check, one can verify that the above identities are consistent with Eq.~\eqref{tr_CP}, i.e.,
    \be
    \bar{\mC}\coloneqq
    \bar{R}^{\alpha\beta\gamma\delta}\bar{R}_{\alpha\beta\gamma\delta}-2\bar{R}^{\alpha\beta}\bar{R}_{\alpha\beta}+\fr{1}{3}\bar{R}^2
    =\Omega^{-2}\mC.
    \label{Weyl2bar}
    \ee

Let us now apply the above formulae to the conformal Kerr metric, where the seed metric~$g_\mn$ is given by the Kerr metric and $\Omega=\Omega(\mP)$ with $\mP$ being the Chern-Simons term associated with the seed Kerr metric.
We assume $\Omega(\mP=0)=1$ so that the conformal factor approaches unity for large $r$.
In this case, the terms with $R^\alpha_\beta$ or $R$ vanish in Eq.~\eqref{curv_inv_conformal}.
Since the full expression is involved, here we consider either the small-$a$ limit or the large-$r$ limit.
Also, we only present the expression for $\bar{R}^{\alpha\beta}\bar{R}_{\alpha\beta}$ and $\bar{R}$ because the Kretschmann scalar~$\bar{R}^{\alpha\beta\gamma\delta}\bar{R}_{\alpha\beta\gamma\delta}$ can be computed from the other two through Eq.~\eqref{Weyl2bar}.
In the small-$a$ limit, we have
    \begin{align}
    \begin{split}
    \bar{R}^{\alpha\beta}\bar{R}_{\alpha\beta}
    &=\frac{165888a^2M^4\Omega'^2[64r(r-2M)+(3168r^2-14600Mr+16905M^2)\cos^2\theta]}{r^{20}}+\mO(a^3), \\
    \bar{R}
    &=\frac{1728aM^2\Omega'(20r-49M)\cos\theta}{r^{10}}+\mO(a^2),
    \end{split}
    \end{align}
where $\Omega'\coloneqq \D\Omega/\D\mP|_{\mP=0}$.
In the large-$r$ limit, we have
    \begin{align}
    \begin{split}
    \bar{R}^{\alpha\beta}\bar{R}_{\alpha\beta}
    &=\frac{5308416a^2M^4\Omega'^2(2+99\cos^2\theta)}{r^{18}}+\mO(r^{-19}), \\
    \bar{R}
    &=\frac{34560aM^2\Omega'\cos\theta}{r^9}+\mO(r^{-10}).
    \end{split}
    \end{align}

%%%%%%%%%%%%%%%%%%%%%%%%%%%%%%%%%%%%%%%%%%
\section{Effective potential for circular orbits}
\label{appC}

In this appendix, we argue that circular orbits of a freely falling test particle on a general stationary axisymmetric background spacetime as well as the stability of the orbits can be described by an effective potential.\footnote{A similar technique has been used in the literature, e.g., Refs.~\cite{Wunsch:2013st,Dolan:2016bxj,Cunha:2017qtt,Zhang:2018qdk,Nakashi:2019mvs}, but their analyses are restricted to a specific spacetime and/or a null test particle.
Our analysis here applies to the general stationary axisymmetric metric~\eqref{genmet}, and the test particle can be either timelike or null.}
Let us consider the most general stationary axisymmetric metric that is invariant under $(t,\varphi)\to (-t,-\varphi)$,
    \be \label{genmet}
    g_\mn\D x^\mu\D x^\nu
    =g_{tt}(r,\theta)\D t^2+g_{rr}(r,\theta)\D r^2+g_{\theta\theta}(r,\theta)\D\theta^2+g_{\varphi\varphi}(r,\theta)\D\varphi^2+2g_{t\varphi}(r,\theta)\D t\D\varphi.
    \ee
Note that there remains a gauge degree of freedom, and one could choose a gauge to reduce the metric~\eqref{genmet} to the so-called Lanczos form~\cite{Lanczos1924}.
However, we do not do so since Eq.~\eqref{genmet} is more useful for the application to the conformal Kerr metric (see \S\ref{sec:timelikegeodesic}).

The motion of a test particle is governed by the following geodesic equation:
    \be
    \ddot{x}^\lambda + \Gamma^\lambda_{\mu\nu} \dot{x}^{\mu} \dot{x}^{\nu}=0, 
    \label{eq:geodesics_appC}
    \ee
where a dot denotes the derivative with respect to an affine parameter~$s$ and $\Gamma^\lambda_{\mu\nu}$ is the Christoffel symbol associated with the metric~$g_\mn$.
Here, we define
    \begin{align}
    -\kappa&\coloneqq g_{\alpha\beta}\dot{x}^\alpha\dot{x}^\beta
    =g_{tt}\dot{t}^2 + 2g_{t\varphi}\dot{t}\dot{\varphi} + g_{\varphi\varphi}\dot{\varphi}^2 + g_{rr} \dot{r}^2 + g_{\theta\theta} \dot{\theta}^2.
    \end{align}
Note that timelike and null geodesics have $\kappa=1$ and $0$, respectively.
As was the case for the (conformal) Kerr metric, there exist two conserved quantities
    \be
    \begin{split}
    -E&\coloneqq g_{t\mu}\dot{x}^\mu
    =g_{tt}\dot{t}+g_{t\varphi}\dot{\varphi}, \\
    L_z&\coloneqq g_{\varphi\mu}\dot{x}^\mu
    =g_{t\varphi}\dot{t}+g_{\varphi\varphi}\dot{\varphi}.
    \end{split}
    \ee

Let us now define an effective potential~$V_{\rm eff}(r,\theta)$ by
    \be
    V_{\rm eff}\coloneqq \kappa+g_{tt}\dot{t}^2 + 2g_{t\varphi}\dot{t}\dot{\varphi} + g_{\varphi\varphi}\dot{\varphi}^2
    =\kappa+\fr{L_z^2 g_{tt}+2EL_z g_{t\varphi}+E^2 g_{\varphi\varphi}}{g_{tt}g_{\varphi\varphi}-g_{t\varphi}^2},
    \ee
which satisfies
    \be
    g_{rr} \dot{r}^2 + g_{\theta\theta} \dot{\theta}^2+V_{\rm eff}=0.
    \label{eq:constraint_appC}
    \ee
In what follows, we show that a circular orbit, defined by $r(s)={\rm const}\eqqcolon r_0$ and $\theta(s)={\rm const}\eqqcolon \theta_0$, can be discussed in terms of the above $V_{\rm eff}$.
First, Eq.~\eqref{eq:constraint_appC} immediately tells us that
    \be
    V_{\rm eff}(r_0,\theta_0)=0.
    \ee
Second, imposing $r(s)=r_0$ and $\theta(s)=\theta_0$ in the $r$- and $\theta$-components of the geodesic equation~\eqref{eq:geodesics_appC} yields
    \be
    \pa_r V_{\rm eff}(r_0,\theta_0)=0, \qquad
    \pa_\theta V_{\rm eff}(r_0,\theta_0)=0,
    \ee
respectively.
Finally, one can also show that the (in)stability of the circular orbit under linear perturbations is encoded in the second derivatives of $V_{\rm eff}$.
In order to investigate the stability, let us consider small fluctuations~$\delta r(s)\coloneqq r(s)-r_0$ and $\delta \theta(s)\coloneqq \theta(s)-\theta_0$ about the circular orbit and study their evolution based on the geodesic equation~\eqref{eq:geodesics_appC}.
At the first order in the fluctuations, we have
    \be
    \fr{\D^2}{\D s^2}
    \bem
    \delta r \\ \delta \theta
    \eem
    =-A
    \bem
    \delta r \\ \delta \theta
    \eem, \qquad
    A\coloneqq \fr{1}{2}\left.
    \bem
    \pa_r^2 V_{\rm eff}/g_{rr} & \pa_r\pa_\theta V_{\rm eff}/g_{rr} \\
    \pa_r\pa_\theta V_{\rm eff}/g_{\theta\theta} & \pa_\theta^2 V_{\rm eff}/g_{\theta\theta}
    \eem\right|_{(r,\theta)=(r_0,\theta_0)}.
    \ee
Therefore, the circular orbit would be stable if both the eigenvalues of the matrix~$A$ is positive, or equivalently, if $\det A>0$ and ${\rm tr}\, A>0$.
Note that the off-diagonal components of $A$ vanish if the background spacetime has an equatorial symmetry ($\pa_\theta g_{\mu\nu}=0$ at $\theta=\pi/2$) and one considers an equatorial circular orbit with $\theta_0=\pi/2$.
In this case, the system of equations for $\delta r$ and $\delta\theta$ is decoupled and our analysis reduces to that in Ref.~\cite{Ono:2016lql}.
In particular, one can show that an equatorial circular orbit in the Kerr spacetime is stable under linear perturbations in both $r$- and $\theta$-directions~\cite{Ono:2016lql}.

To summarize, a circular orbit with $r(s)=r_0$ and $\theta(s)=\theta_0$ satisfies
    \be
    V_{\rm eff}(r_0,\theta_0)=0, \qquad
    \pa_r V_{\rm eff}(r_0,\theta_0)=0, \qquad
    \pa_\theta V_{\rm eff}(r_0,\theta_0)=0,
    \ee
and it would be stable under linear perturbations if
    \be
    \pa_r^2 V_{\rm eff}\,\pa_\theta^2 V_{\rm eff}-(\pa_r\pa_\theta V_{\rm eff})^2>0 \quad
    \text{and} \quad
    \fr{1}{g_{rr}}\pa_r^2 V_{\rm eff}+\fr{1}{g_{\theta\theta}}\pa_\theta^2 V_{\rm eff}>0 \quad
    \text{at $(r,\theta)=(r_0,\theta_0)$},
    \ee
where we have assumed that $g_{rr}>0$ and $g_{\theta\theta}>0$.

%%%%%%%%%%%%%%%%%%%%%%%%%%%%%%%%%%%%%%%%%%%%%%%%%%%%%%%%%%%%%%%%%%%%%%%%%%%%%%%%%%%%
%%%%%%%%%%%%%%%%%%%%%%%%%%%%%%%%%%%%%%%%%%%%%%%%%%%%%%%%%%%%%%%%%%%%%%%%%%%%%%%%%%%%

\bibliographystyle{mybibstyle}
\bibliography{bib}

\end{document}